\newif\ifarxiv
\arxivtrue
\pdfoutput=1
\documentclass[a4paper,final,11pt]{siamart220329}
\usepackage[l2tabu, orthodox]{nag}

\usepackage{amsmath}
\usepackage{amssymb}
\usepackage[subrefformat=parens,labelformat=parens,labelfont={}]{subcaption}
\usepackage{sidecap}
\ifarxiv
  \usepackage[frozencache=true,cachedir=_minted-paper]{minted}
\else
  \usepackage[finalizecache=true,cachedir=_minted-paper]{minted}
\fi
\usepackage{hyperref}
\usepackage[margin=2.5cm]{geometry}
\usepackage{nomencl}
\usepackage{booktabs}
\usepackage{multirow}
\usepackage{microtype}
\makenomenclature
\usepackage{
tikz,
pgfplots,
pgfplotstable,
ifthen,
}
\usetikzlibrary{fit}
\usetikzlibrary{calc}
\usetikzlibrary{shapes.geometric}
\usetikzlibrary{backgrounds}
\pgfplotsset{compat=1.15}

\newcommand{\objname}[1]{\texttt{#1}}
\newcommand{\minus}{\scalebox{0.75}[1.0]{\hspace{2pt}\( - \)\hspace{2pt}}}

\newcommand{\aset}[1]{{#1}}

\newcommand{\anarray}[1]{\textsf{#1}}

\DeclareRobustCommand*{\ol}{\overleftarrow}
 
\newcommand{\tikzscale}{10}
\pgfmathsetmacro{\tikzsqrtthree}{1.7320508075688772}
\pgfmathsetmacro{\tikzsqrttwo}{1.414213}
\pgfmathsetmacro{\tikzoffx}{.85}
\pgfmathsetmacro{\tikzoffy}{1}
\pgfmathsetmacro{\tikzscale}{0.65}
\pgfmathsetmacro{\tikzoffmeshB}{1.5}

\definecolor{amber}{rgb}{1.0, 0.75, 0.0}
\definecolor{armygreen}{rgb}{0.29, 0.33, 0.13}
\definecolor{azure(colorwheel)}{rgb}{0.0, 0.5, 1.0}
\definecolor{awesome}{rgb}{1.0, 0.13, 0.32}
\newcommand{\tikzvarlen}{0.4}

\pgfmathsetmacro{\tikzpictureoffset}{6.0}
\newcommand{\tikzmeshcelllen}{2.25}
\newcommand{\tikzveccelllen}{0.5}
\newcommand{\tikzrankoff}{5}

\title{Efficient N-to-M Checkpointing Algorithm for Finite Element Simulations}
\author{%
  David A.~Ham\thanks{Department of Mathematics at Imperial College London (\email{david.ham@imperial.ac.uk})}
  \and
  Vaclav Hapla\thanks{Department of Earth and Planetary Sciences at ETH Zurich \& Department of Applied Mathematics, FEECS at VSB-TU Ostrava (\email{vaclav.hapla@gmail.com})}
  \and
  Matthew G.~Knepley\thanks{Department of Computer Science and Engineering at the University at Buffalo (\email{knepley@gmail.com})}
  \and
  Lawrence Mitchell\thanks{NVIDIA Corporation, Santa Clara (\email{lmitchell@nvidia.com})}
  \and
  Koki Sagiyama\thanks{Department of Mathematics at Imperial College London (\email{k.sagiyama@imperial.ac.uk})}
}

\usepackage{cleveref}
\headers{Efficient N-to-M Checkpointing for FE Simulations}{D.~A.~Ham, V.~Hapla, M.~G.~Knepley, L.~Mitchell, and K.~Sagiyama}
\begin{document}
\maketitle

\begin{abstract}
In this work, we introduce a new algorithm for N-to-M checkpointing in finite element simulations.
This new algorithm allows efficient saving/loading of functions representing physical quantities associated with the mesh representing the physical domain.
Specifically, the algorithm allows for using different numbers of parallel processes for saving and loading, allowing for restarting and post-processing on the process count appropriate to the given phase of the simulation and other conditions.
For demonstration, we implemented this algorithm in PETSc, the Portable, Extensible Toolkit for Scientific Computation, and added a convenient high-level interface into Firedrake, a system for solving partial differential equations using finite element methods.
We evaluated our new implementation by saving and loading data involving 8.2 billion finite element degrees of freedom using 8,192 parallel processes on ARCHER2, the UK National Supercomputing Service.
\end{abstract}

\ifarxiv
\else
\fi

\section{Introduction}\label{S:introduction}
In many fields of science and engineering, there has been an increasing demand for large-scale simulations of complex physical problems using finite element methods run on distributed-memory computer clusters and supercomputers.
When running a simulation on such computing systems,
it is crucial for the finite element analysis software package to have scalable and flexible checkpointing that allows for saving a state of the simulation, i.e., meshes representing the computational domain and functions representing the physical quantities of interest, to disk at an arbitrary time and loading it later using a process count appropriate to the given phase of the simulation.
Such a capability enables users to complete simulations in multiple sessions since computing systems often enforce a maximum allowed run time on each submitted job.
Moreover, long-running applications can sometimes be unexpectedly terminated for various reasons such as running out of wall time or system failure.

The ability to use different process counts for saving and loading is crucial, e.g., when the user desires to run a simulation on a supercomputer using a large number of parallel processes and later post-process the result on a local workstation using a much smaller number of processes.
Algorithms with this flexibility are sometimes called N-to-M checkpointing algorithms, as they allow saving from N processes and loading on M processes,
as opposed to N-to-N algorithm, which requires that the saving and loading process counts match.
To our knowledge, major finite element problem-solving environments, such as DOLFIN/DOLFINx from the FEniCS Project~\cite{fenics}, MOOSE~\cite{moose} when using advanced restarting, and deal.II~\cite{dealii} when using the general fully distributed meshes, have only supported N-to-N checkpointing, and Firedrake~\cite{Rathgeber2016,FiredrakeUserManual} hitherto also had only had N-to-N checkpointing.\@ deal.II, when using a special mesh class that requires a shared coarse mesh across parallel processes, supports N-to-M checkpointing via p4est~\cite{p4est}, a software library for adaptive mesh refinement on forests of octrees. This feature requires the user to provide a shared coarse mesh object on each rank when loading that represents the same coarse mesh as used at the time of saving. This coarse mesh allows for reconstructing functions that have been saved as cell-localised data on the loaded mesh.

A mesh is represented by entities, i.e., cells, faces, edges, and vertices, and their connectivities. We define finite element function spaces on the mesh, with each degree of freedom (DoF) associated with some mesh entity. Possibly, we can associate multiple DoFs with the same mesh entity. A function in a finite element function space is represented by a vector
containing the DoF values. The main challenge in developing a scalable and efficient N-to-M checkpointing algorithm that works with general fully distributed meshes stems from the fact that, on load, one needs to handle the computational mesh arbitrarily redistributed over the participating processes, irrespective of how the mesh was distributed at the time of saving, while keeping track of the mapping of the saved DoF values to members of the dual space defined on the loaded mesh.

While the association of DoF values in the vector with entities in the mesh is effectively preserved when we save and later load the function, 
the mesh might be distributed completely differently, possibly over a different number of processes, so
we need to re-establish which entity on the loaded mesh a given DoF value is defined on and which DoF on that entity the DoF value is associated with.
For the former, one needs to establish an entity-entity association between the mesh before saving and the loaded mesh.
For the latter, one needs a method to \emph{order} function space DoFs in a deterministic manner on each entity.

In this work, we develop an efficient N-to-M checkpointing algorithm and implement it in Firedrake and the Portable, Extensible Toolkit for Scientific Computation (PETSc)~\cite{petsc-web-page, petsc-user-ref, petsc-efficient}.
In~\cite{HaplaKnepleyAfanasievBoehmDrielKrischerFichtner2020},
fully parallel N-to-M checkpointing was introduced for meshes using the XDMF file format, which only required saving and loading cell-vertex connectivity data, and faces and/or edges were recovered algorithmically from the cell-vertex connectivity after loading.
Though that algorithm is efficient, it is not suited for saving/loading function spaces as it cannot store function space data associated with faces and edges.
We also found that we do not benefit from the XDMF metadata file format due to insufficient support for the wide range of finite elements that Firedrake supports.
Thus, in this work, we implement our N-to-M checkpointing algorithm using a new PETSc-specific format based on HDF5~\cite{hdf5}.
As Firedrake meshes, function spaces, and functions are almost directly represented by corresponding PETSc objects, we first implement relevant functions in PETSc and then wrap them in Firedrake.
Hence, packages depending on Firedrake or PETSc can directly benefit from this work as well.

In \cref{S:concept} we explain the concept of meshes and functions checkpointing.
Specifically, we deal with saving and loading meshes in \cref{SS:mesh} and 
saving and loading function spaces and functions in \cref{SS:function}.
We then describe how we broadcast the loaded function space and function data onto the separately loaded mesh in \cref{SS:reconstruct}.
Then, in \cref{S:impl}, we translate the concepts introduced in \cref{S:concept} to a scalable implementation in PETSc.
\Cref{S:orientations} describes treatment of \emph{orientations} in Firedrake necessary to incorporate the implementation introduced in \cref{S:impl}.
In \cref{S:firedrake_api} we illustrate example usages of our new checkpointing feature.
In \cref{S:evaluation} we perform weak scalability tests for our implementations on ARCHER2.
Finally, \cref{S:conclusion} summarises this work.
We provide a list of the symbols used in this paper in \cref{S:symbols}.

\section{Saving and loading meshes, function spaces, and functions}\label{S:concept}
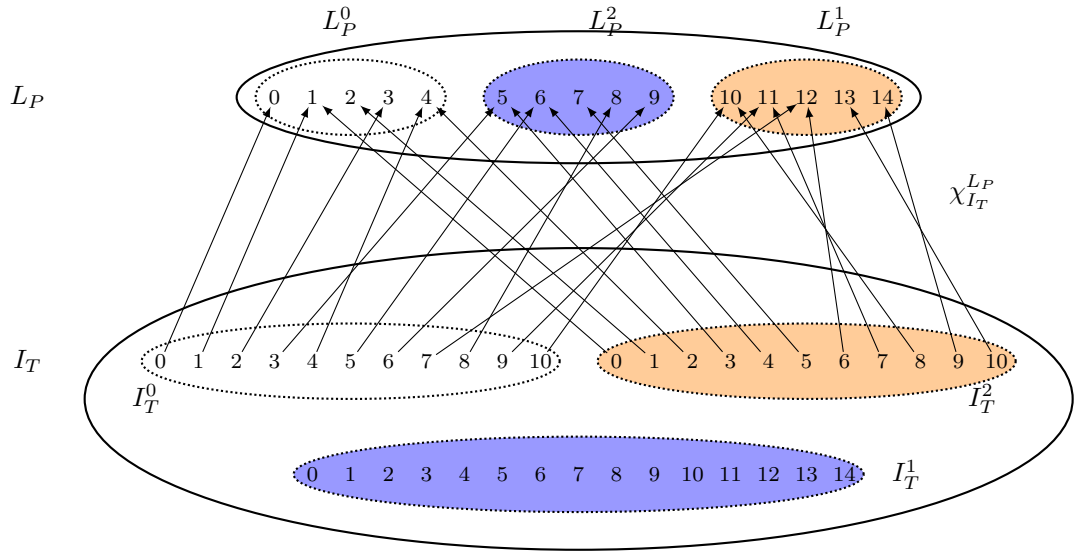
\begin{figure}[ht]
\begin{tikzpicture}[darkstyle/.style={circle,draw,fill=gray!40}]
  \tikzstyle{setp}=[black, thick]
  \tikzstyle{subsetp0}=[black,thick,densely dotted]
  \tikzstyle{subsetp1}=[black,thick,densely dotted,fill=blue!40]
  \tikzstyle{subsetp2}=[black,thick,densely dotted,fill=orange!40]

  \node[inner sep=1pt] (XOO) at (0,0) {};
  \node[inner sep=1pt] (XO) at ($(XOO)+(\tikzveccelllen*5, 0)$) {};
  \node[inner sep=1pt] () at ($(XOO)+\tikzveccelllen*(-1,.5)$) {$\aset{L}_P$};
  \foreach \x in {0,1,...,14}
  {
    \ifthenelse{\x=0 \OR \x=1 \OR \x=2 \OR \x=3 \OR \x=4}
    {\node[inner sep=1pt] (XBL\x) at ($(XO)+\tikzveccelllen*(\x,0)$) {};} {}
    \ifthenelse{\x=5 \OR \x=6 \OR \x=7 \OR \x=8 \OR \x=9}
    {\node[inner sep=1pt] (XBL\x) at ($(XO)+\tikzveccelllen*(\x+1,0)$) {};} {}
    \ifthenelse{\x=10 \OR \x=11 \OR \x=12 \OR \x=13 \OR \x=14}
    {\node[inner sep=1pt] (XBL\x) at ($(XO)+\tikzveccelllen*(\x+2,0)$) {};} {}
    \node[inner sep=1pt] (XTR\x) at ($(XBL\x)+\tikzveccelllen*(1,1)$) {};
    \node (XT\x) at ($(XTR\x)-\tikzveccelllen*(.5,0)$) {};
    \node (XB\x) at ($(XBL\x)+\tikzveccelllen*(.5,0)$) {};
    \node (X\x) at ($(XBL\x)!0.5!(XTR\x)$) {};
  }
  \node[inner sep=1pt] (COO) at ($(XOO)+(0, -\tikzveccelllen*7)$) {};
  \node[inner sep=1pt] (CO) at ($(COO)+\tikzveccelllen*(2, 0)$) {};
  \node[inner sep=1pt] (CLabel) at ($(COO)+\tikzveccelllen*(-1,.5)$) {$\aset{I}_T$};
  \foreach \x in {0,1,...,10}
  {
    \node[inner sep=1pt] (CBL\x) at ($(CO)+(\tikzveccelllen*\x, 0)$) {};
    \node[inner sep=1pt] (CTR\x) at ($(CO)+(\tikzveccelllen*\x+\tikzveccelllen, \tikzveccelllen)$) {};
    \node (C\x) at ($(CBL\x)!0.5!(CTR\x)$) {};
    \node (CT\x) at ($(CO)+(\tikzveccelllen*\x+\tikzveccelllen/2, \tikzveccelllen)$) {};
    \node (CB\x) at ($(CO)+(\tikzveccelllen*\x+\tikzveccelllen/2, 0)$) {};
  }
  \node[inner sep=1pt] (DOO) at ($(COO)+\tikzveccelllen*(4, -3)$) {};
  \node[inner sep=1pt] (DO) at ($(DOO)+\tikzveccelllen*(2, 0)$) {};
  \foreach \x in {0,1,...,14}
  {
    \node[inner sep=1pt] (DBL\x) at ($(DO)+(\tikzveccelllen*\x, 0)$) {};
    \node[inner sep=1pt] (DTR\x) at ($(DO)+(\tikzveccelllen*\x+\tikzveccelllen, \tikzveccelllen)$) {};
    \node (D\x) at ($(DBL\x)!0.5!(DTR\x)$) {};
    \node (DT\x) at ($(DO)+(\tikzveccelllen*\x+\tikzveccelllen/2, \tikzveccelllen)$) {};
    \node (DB\x) at ($(DO)+(\tikzveccelllen*\x+\tikzveccelllen/2, 0)$) {};
  }
  \node[inner sep=1pt] (EOO) at ($(COO)$) {};
  \node[inner sep=1pt] (EO) at ($(EOO)+\tikzveccelllen*(14, 0)$) {};
  \foreach \x in {0,1,...,10}
  {
    \node[inner sep=1pt] (EBL\x) at ($(EO)+(\tikzveccelllen*\x, 0)$) {};
    \node[inner sep=1pt] (ETR\x) at ($(EO)+(\tikzveccelllen*\x+\tikzveccelllen, \tikzveccelllen)$) {};
    \node (E\x) at ($(EBL\x)!0.5!(ETR\x)$) {};
    \node (ET\x) at ($(EO)+(\tikzveccelllen*\x+\tikzveccelllen/2, \tikzveccelllen)$) {};
    \node (EB\x) at ($(EO)+(\tikzveccelllen*\x+\tikzveccelllen/2, 0)$) {};
  }
  \draw [subsetp0] ($(X2)$) ellipse [x radius=\tikzveccelllen*2.5, y radius=\tikzveccelllen*1];
  \draw [subsetp1] ($(X7)$) ellipse [x radius=\tikzveccelllen*2.5, y radius=\tikzveccelllen*1];
  \draw [subsetp2] ($(X12)$) ellipse [x radius=\tikzveccelllen*2.5, y radius=\tikzveccelllen*1];
  \draw [subsetp0] ($(C5)$) ellipse [x radius=\tikzveccelllen*5.5, y radius=\tikzveccelllen*1];
  \draw [subsetp1] ($(D7)$) ellipse [x radius=\tikzveccelllen*7.5, y radius=\tikzveccelllen*1];
  \draw [subsetp2] ($(E5)$) ellipse [x radius=\tikzveccelllen*5.5, y radius=\tikzveccelllen*1];
  \draw [setp] ($(X7)-\tikzveccelllen*(0,0)$) ellipse [x radius=\tikzveccelllen*9, y radius=\tikzveccelllen*1.75];
  \draw [setp] ($(C0)+\tikzveccelllen*(11,-1)$) ellipse [x radius=\tikzveccelllen*13, y radius=\tikzveccelllen*4];
  \path[latex-, draw=black] (X0)--(C0);
  \path[latex-, draw=black] (X1)--(C1);
  \path[latex-, draw=black] (X3)--(C2);
  \path[latex-, draw=black] (X5)--(C3);
  \path[latex-, draw=black] (X4)--(C4);
  \path[latex-, draw=black] (X6)--(C5);
  \path[latex-, draw=black] (X9)--(C6);
  \path[latex-, draw=black] (X12)--(C7);
  \path[latex-, draw=black] (X8)--(C8);
  \path[latex-, draw=black] (X11)--(C9);
  \path[latex-, draw=black] (X10)--(C10);
  \path[latex-, draw=black] (X1)--(E0);
  \path[latex-, draw=black] (X2)--(E1);
  \path[latex-, draw=black] (X4)--(E2);
  \path[latex-, draw=black] (X5)--(E3);
  \path[latex-, draw=black] (X6)--(E4);
  \path[latex-, draw=black] (X7)--(E5);
  \path[latex-, draw=black] (X12)--(E6);
  \path[latex-, draw=black] (X11)--(E7);
  \path[latex-, draw=black] (X10)--(E8);
  \path[latex-, draw=black] (X14)--(E9);
  \path[latex-, draw=black] (X13)--(E10);
  \foreach \x in {0,1,...,14}
  {
    \node () at ($(X\x)$) {\footnotesize \x};
  }
  \foreach \x in {0,1,...,10}
  {
    \node () at ($(C\x)$) {\footnotesize \x};
  }
  \foreach \x in {0,1,...,14}
  {
    \node () at ($(D\x)$) {\footnotesize \x};
  }
  \foreach \x in {0,1,...,10}
  {
    \node () at ($(E\x)$) {\footnotesize \x};
  }
  \node[anchor=west] () at ($(XOO)+(\tikzveccelllen*23,-\tikzveccelllen*2)$) {$\chi_{\aset{I}_T}^{\aset{L}_P}$};
  \node[anchor=west] () at ($(C0)+\tikzveccelllen*(-1,-1)$) {$\aset{I}_T^{0}$};
  \node[anchor=west] () at ($(E9)+\tikzveccelllen*(0,-1)$) {$\aset{I}_T^{2}$};
  \node[anchor=west] () at ($(E7)+\tikzveccelllen*(0,-3)$) {$\aset{I}_T^{1}$};
  \node[anchor=west] () at ($(X2)+\tikzveccelllen*(-1,2)$) {$\aset{L}_P^{0}$};
  \node[anchor=west] () at ($(X7)+\tikzveccelllen*(0,2)$) {$\aset{L}_P^{2}$};
  \node[anchor=west] () at ($(X12)+\tikzveccelllen*(0,2)$) {$\aset{L}_P^{1}$};
\end{tikzpicture}
\caption{
A schematic of the map $\chi_{\aset{I}_T}^{\aset{L}_P}$ from the local numbers of the loaded mesh topology, $\aset{I}_T$, to the global numbers, $\aset{L}_P$,
for the example depicted in \cref{Fi:mesh},
specifically \cref{Fi:mesh_loaded_num}.
Parts owned by processes 0, 1, and 2 are shown in white, blue, and orange, respectively.
Only maps from $\aset{I}_T^{(0)}$ and $\aset{I}_T^{(2)}$ to $\aset{L}_P$ are depicted for simplicity.
This map is to be composed with the inverse of $\chi_{\aset{I}_P}^{\aset{L}_P}$ depicted in \cref{Fi:function_map}.
\label{Fi:mesh_map}}
\end{figure}

We save mesh, function space, and function data from $N$ processes, and later
load those data and create new parallel representations of the mesh, function space, and function on $M$ processes.
Loading data and reconstructing the mesh and the function space/function require
many maps and map compositions,
which are the key components of the algorithm that we introduce in this section.
We use an example finite element problem throughout this section.
For this example problem, those key maps are depicted in
\cref{Fi:function_data_local_loaded} appearing in \cref{SSS:loaded_global_function_space} and \cref{SSS:loaded_global_function},
\cref{Fi:function_map} appearing in \cref{SSS:loaded_global_function_space},
\cref{Fi:mesh_map} appearing in \cref{SSS:loaded_mesh}, and
\cref{Fi:function_data_loaded} appearing in \cref{SSS:loaded_local_function_space,SSS:loaded_local_function},
with arrays viewed as maps from their indices to the array values.
Then, the composition of the maps precisely describes our loading mechanism.
Our goal in this section is to describe the underlying concept of these figures.

\subsection{Meshes}\label{SS:mesh}
\input{paper_fig_save_load_mesh}
We consider a mesh representing a physical domain on which functions such as velocity and pressure are defined.
The mesh is viewed as composed of a topology and coordinates,
but in this section, we only deal with saving and loading the mesh topology;
the coordinates are handled just like other functions;
see \cref{SS:function,SS:reconstruct}.
Note also that in this work, we deal only with ``fully interpolated meshes'',
where all entities of all dimensions are explicitly represented.

The mesh topology is composed of \emph{entities},
such as cells, faces, edges, and vertices,
and is described by their connectivities.
We denote by $E$ the total number of entities in the mesh topology.
We assign a unique non-negative integer, called a \emph{global number}
and drawn from a set
\begin{align}
\aset{I}:=\{0,\dots,E\minus 1\},
\end{align}
to each entity.

Crucial in this work is the concept of a \emph{cone}.
The \emph{cone} of an $d$-dimensional entity is defined to be
an \emph{ordered} list of the $(d-1)$-dimensional entities
directly attached to that entity.
For instance, a two-dimensional triangular cell has three one-dimensional edges attached.
The cone of that triangular cell is then an ordered list of the edges;
the cone of a vertex is an empty list.
This representation of meshes is detailed in~\cite{LangeMitchellKnepleyGorman2015}.

\Cref{Fi:mesh_global} shows an example mesh topology
with entities labeled with their global numbers.
In this figure, the orderings of the cones are indicated by the arrows.

\begin{figure}[ht]
\hspace{50pt}
\begin{tikzpicture}
    \tikzstyle{point}=[circle,fill,minimum size=5pt,inner sep=0.5pt]
    \tikzstyle{dof}=[circle,fill,inner sep=1.5pt]
    \tikzstyle{topoedge}=[draw=black,thick,shorten <= 4pt,shorten >= 4pt,-latex]
    \tikzstyle{fsnode}=[circle,draw=black,minimum size=11pt,inner sep=0pt,fill=white]
    \node (v3) [inner sep=0pt] at ($\tikzmeshcelllen*(1,0)$) {};
    \node (v4) [inner sep=0pt] at ($\tikzmeshcelllen*(0,0)$) {};
    \node[point] (p6) at ($\tikzmeshcelllen*(-2, 0)+(v3)$) {};
    \node[point] (p7) at ($\tikzmeshcelllen*(-2, 0)+(v4)$) {};
    \path[topoedge] ($(p6)$)--($(p7)$);
    \node[point] (p6) at ($\tikzmeshcelllen*(0, 0)+(v3)$) {};
    \node[point] (p7) at ($\tikzmeshcelllen*(0, 0)+(v4)$) {};
    \path[topoedge] ($(p6)$)--($(p7)$);
    \node[fsnode] (fsnode3) at ($(p6)!0.25!(p7)$) {\footnotesize $\sigma_0$};
    \node[fsnode] (fsnode4) at ($(p6)!0.50!(p7)$) {\footnotesize $\sigma_1$};
    \node[fsnode] (fsnode5) at ($(p6)!0.75!(p7)$) {\footnotesize $\sigma_2$};
    \node[point] (p6) at ($\tikzmeshcelllen*(2, 0)+(v3)$) {};
    \node[point] (p7) at ($\tikzmeshcelllen*(2, 0)+(v4)$) {};
    \path[topoedge] ($(p6)$)--($(p7)$);
    \node[fsnode] (fsnode3) at ($(p6)!0.25!(p7)$) {\footnotesize $\sigma_2$};
    \node[fsnode] (fsnode4) at ($(p6)!0.50!(p7)$) {\footnotesize $\sigma_1$};
    \node[fsnode] (fsnode5) at ($(p6)!0.75!(p7)$) {\footnotesize $\sigma_0$};
\end{tikzpicture}
\caption{
Example illustrating the cone ordering and the DoF ordering of a one-dimensional one-element mesh.
(Left) A one-dimensional one-element mesh with the cone of the cell entity ordered so that the right vertex comes first and then the left vertex, which is indicated by the arrow.
(Middle) A schematic of the DP2 function space with three DoFs, $\sigma_0$, $\sigma_1$, and $\sigma_2$, ordered in the cell in a deterministic way based on the cone ordering.
(Right) A schematic of the DP2 function space with DoFs wrongly ordered according to our definition. The DoF values in the DoF vector will be associated with the wrong DoFs.
\label{Fi:orientation_1d}}
\end{figure}
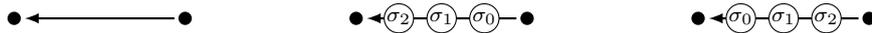

When we save the mesh topology, we save the cone of each entity directly,
so that, after loading the mesh topology,
that entity can have the same cone ordering as before saving.
It is crucial for the orderings of the cones to be preserved in the save-load cycle
in order to save and load a function defined on the mesh.
If multiple DoFs are associated with the same mesh entity,
we need to \emph{order} them in a certain way to order the associated DoF values in the DoF vector representing the function.
When we save and load the DoF vector, the ordering of the DoF values will be unchanged.
To reestablish the association of those DoF values with the correct DoFs on the loaded mesh,
which might be arbitrarily distributed across the loading processes,
we need to be able to order the DoFs on the interested mesh entity in the exact same way as before saving, or the DoF values might end up being associated with the wrong DoFs.
The ordering of the cone of that mesh entity can be used, as they are preserved in the save-load cycle, to order DoFs on that entity in a deterministic manner.
\Cref{Fi:orientation_1d} shows a simple one-dimensional example that exhibits the idea.

\subsubsection{Saving mesh topology}\label{SSS:saved_mesh}
Suppose that the mesh is distributed over
parallel processes so that
only a part of the mesh is visible to a given process.
An entity may be visible to multiple processes,
but one of those processes is to have the \emph{ownership}
of that entity and the other processes are to view it as a \emph{ghost} (not owned) entity.
The ordering of the cone of each entity is to be preserved upon mesh distribution.
Here and thereafter, we use $\ol{\cdot}$ for sets, arrays, and other quantities before saving
to distinguish them from those after loading.
We use subscript $T$ for sets, arrays, and quantities associated with the mesh topology,
$T$ standing for \emph{topology}.

If $\ol{E}_T^{n}$ is the total number of entities on the mesh topology visible to
process $n\in\{0,\dots,N \minus 1\}$,
where $N$ is the number of processes used in the saving session,
each entity on process $n$ is labeled with a unique \emph{local number}
drawn from the set
\begin{align}
\ol{\aset{I}}_T^{n}:=\{0,\dots,\ol{E}_T^{n}\minus 1\}.
\label{E:ITn}
\end{align}
We keep a mapping associating the local number of each entity with its global number.
Thus, the entity's cone can be written in both global and local numbering.
When the mesh topology is saved, each process saves the cones of its owned entities
in terms of global numbers.

\Cref{Fi:mesh_saved} shows an example distribution of the mesh topology shown in
\cref{Fi:mesh_global},
where, on each process, entities are labeled with local numbers.
Note that the cones are preserved upon distribution.
In this example, on the process $0$ mesh topology, for instance,
local number $0$ is associated with global number $1$ in the original mesh topology
shown in \cref{Fi:mesh_global},
and the cone of this cell entity is
$(6,7,8)$ in terms of local numbers and
$(12,11,10)$ in terms of global numbers.
When saving the mesh topology,
process 0 and process 1 save the cones of their owned entities
in global numbering.

\subsubsection{Loading mesh topology}\label{SSS:loaded_mesh}
We now consider loading the saved mesh topology in parallel;
we denote by $M$ the number of loading processes.
In general, the number of processes used for loading, $M$, is
different from that used for saving, $N$.
The saved cone data, stored in global numbering, are loaded onto $M$ processes,
which then reconstruct the mesh topology.
This is, in general, a multi-step process; the loading processes first
construct a mesh topology with no cell overlaps using a naive partition of the saved data, then
redistribute it using a mesh partitioner, such as ParMETIS~\cite{parmetis}, to minimise the cost of parallel communication, and finally
add overlaps across process boundaries if necessary.
The details are given in \cref{S:loaded_mesh_details}, and, hereafter, we focus on the final product, which we call the \emph{loaded mesh topology}.
In this section we construct one important map $\chi_{\aset{I}_T}^{\aset{L}_P}$ \cref{E:chi_IT_LP},
depicted in \cref{Fi:mesh_map} for our example,
which associates entities on the loaded parallel mesh topology with the global numbers $\aset{I}$.
Here and elsewhere, we use subscript $P$ for the partitioned sets, arrays, and quantities,
$P$ standing for \emph{partitioned}.

Let $E_T^{m}$ denote the number of entities visible to process $m\in \{0,\dots,M\minus 1\}$
on the loaded mesh topology.
Each entity of the local mesh topology
is arbitrarily labeled with a local number
drawn from set $\aset{I}_T^{m}$,
\begin{align}
\aset{I}_T^{m}:=\{0,\dots,E_T^{m}\minus 1\}.
\label{E:ITm}
\end{align}
Each local number can then be associated with a global number
as the saved cones are represented in terms of global numbers;
we denote by $\anarray{LocG}_T^{m}$ a local array of those global numbers
indexed by $i_T^{m}\in\aset{I}_T^{m}$ so that
$\anarray{LocG}_T^{m}$ represents the map from $\aset{I}_T^{m}$
to the set of all global numbers $\aset{I}$.
To represent these maps monolithically over all parallel processes,
we first define a union set, $\aset{I}_T$, as
\begin{align}
\aset{I}_T:=
\bigcup\limits_{m\in\{0,\dots,M\minus 1\}}\{m\}\times\aset{I}_T^{m},
\label{E:IT}
\end{align}
where $\{.\}\times\{.\}$ represents the Cartesian product of two sets.
Each element $(m,i_T^{m})\in\aset{I}_T$ uniquely identifies
the entity on process $m$ with local number $i_T^{m}\in\aset{I}_T^{m}$.
As mentioned above, by construction,
there is a map from $\aset{I}_T$ to the set of global numbers,
$\aset{I}$, but
it is more consistent and convenient to also partition $\aset{I}$
into $M$ parts, and
index them locally on each process.
We denote by $\aset{L}_P^{m}$ the set of local indices on process $m\in \{0,\dots,M\minus 1\}$,
and define a union set $\aset{L}_P$ as
\begin{align}
\aset{L}_P:=
\bigcup\limits_{m\in\{0,\dots,M\minus 1\}}\{m\}\times\aset{L}_P^{m},
\label{E:LP}
\end{align}
so that each element $(m,l_p^{m})\in\aset{L}_P$ is identified as a unique global number.
A simple bijective map
\begin{align}
\chi_{\aset{I}}^{\aset{L}_P}:
\aset{I}\rightarrow\aset{L}_P,
\label{E:chi_E_LP}
\end{align}
represents the partition.
Thus, with $\aset{I}_T$ \cref{E:IT} and $\aset{L}_P$ \cref{E:LP} defined as in the above,
one can construct a surjective map
\begin{align}
\chi_{\aset{I}_T}^{\aset{L}_P}:\aset{I}_T\rightarrow\aset{L}_P,
\label{E:chi_IT_LP}
\end{align}
upon loading the mesh topology.
The map $\chi_{\aset{I}_T}^{\aset{L}_P}$ is surjective
but, in general, not injective,
as some entities are visible to multiple processes.

With $\aset{I}_T$ and $\aset{L}_P$ defined as in \cref{E:IT} and \cref{E:LP},
we can see $\chi_{\aset{I}_T}^{\aset{L}_P}$ as a \emph{star forest};
a \emph{star}, in graph theory, is a tree with one root node connected to zero or more leaves or an isolated leaf node not connected to any root node, and a star forest is defined to be a collection of stars~\cite{PetscSF2022}.
Here and elsewhere in this work, isolated leaf nodes are not of interest, so we can view a star forest as a map from the set of all leaves to the set of all roots.
In this work, a root is a local number or a local index on some process and
an associated leaf is a local number or a local index on any process (potentially the same)
that is identified with the root.
For $\chi_{\aset{I}_T}^{\aset{L}_P}$,
$\aset{L}_P$ and $\aset{I}_T$ are regarded as the collection of the roots and leaves, respectively.
The concept of star forests is convenient to describe a parallel communication pattern~\cite{PetscSF2022},
and this interpretation is important for a scalable and efficient implementation of our algorithm
as seen in \cref{S:impl}.
Note also that, on each process,
cones are often represented using local numbers, but
the order of the lower-dimensional entities appearing
in the representation of each cone is preserved in the save-load cycle.

\Cref{Fi:mesh_loaded} shows an example mesh topology loaded using three processes.
The process boundary overlaps were added on each process to include a single layer of neighboring cells and the lower dimensional entities directly attached to them.
We observe that the order of the lower-dimensional entities
in the cone representation is preserved in the save-load cycle
by comparing with \cref{Fi:mesh_saved}.
\Cref{Fi:mesh_loaded_num} shows
the sets of local numbers $\aset{I}_T^{m}$ and
the associated arrays of global numbers $\anarray{LocG}_T^{m}$ for $m\in\{0,1,2\}$.
The map $\chi_{\aset{I}_T}^{\aset{L}_P}$ is depicted in \cref{Fi:mesh_map}.

\subsection{Function spaces and functions}\label{SS:function}
We now consider finite element function spaces
defined on the mesh topology that we worked with in \cref{SS:mesh}
and functions defined on those spaces.
A function on a finite element function space is represented by
the values of the degrees of freedom (DoFs),
which are stored in a DoF vector.
We here introduce a notion of \emph{discrete function space data},
which defines the number of DoFs on each relevant entity
of the mesh topology and
the offset of each DoF in the DoF vector.

The mesh coordinates are then fully defined by
the finite element representing the coordinate space and
the corresponding discrete function space data,
as well as the coordinate vector.
Once the mesh coordinates become available,
other functions are fully defined by
the finite elements representing the function spaces and
the corresponding discrete function space data and DoF vectors.
Although the mesh coordinates must first be defined
for other functions to be well-defined,
the procedures to save/load the mesh coordinates and other functions are identical.
As saving and loading the representation of a finite element,
such as family and degree,
are straightforward, we focus here merely on saving and loading
the discrete function space data and DoF vectors.

As mentioned in \cref{SS:mesh}, it is important to save
discrete function space data in terms of global numbers as we did for the mesh topology.
By doing so, even if the mesh and the function space data are later loaded separately, we can reestablish their association by comparing the global numbers.

\input{paper_fig_saved_function}
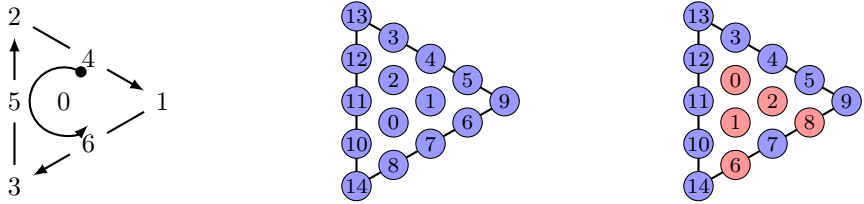
\begin{figure}
\hspace{50pt}
\begin{tikzpicture}
    \tikzstyle{point}=[circle, minimum size=15pt,inner sep=0pt]
    \tikzstyle{point1}=[circle, minimum size=15pt,inner sep=0pt,fill=blue!40]
    \tikzstyle{orntarc}=[x radius=\tikzmeshcelllen*.2, y radius=\tikzmeshcelllen*.2]
    \tikzstyle{dof}=[circle,fill,inner sep=1.5pt]
    \tikzstyle{topoedge0}=[draw=black,thick,shorten <= 8pt,shorten >= 8pt]
    \tikzstyle{topoedge1}=[draw=black,thick,shorten <= 8pt,shorten >= 8pt,-latex]
    \node (v0) [inner sep=0pt] at ($(0, 0)$) {};
    \node (v1) [inner sep=0pt] at ($\tikzmeshcelllen*(0, 1)$) {};
    \node (v2) [inner sep=0pt] at ($\tikzmeshcelllen*(1, 0)$) {};
    \node (v3) [inner sep=0pt] at ($\tikzmeshcelllen*(1, 1)$) {};
    \node (v4) [inner sep=0pt] at ($\tikzmeshcelllen*(1+\tikzsqrtthree/2, .5)$) {};
    \node[point] (p2) at ($\tikzmeshcelllen*(-2, 0)+\tikzmeshcelllen*(1+.5/\tikzsqrtthree, .5)$) {0};
    \node[point] (p5) at ($\tikzmeshcelllen*(-2, 0)+(v2)$) {3};
    \node[point] (p6) at ($\tikzmeshcelllen*(-2, 0)+(v3)$) {2};
    \node[point] (p7) at ($\tikzmeshcelllen*(-2, 0)+(v4)$) {1};
    \node[point] (p11) at ($(p5)!0.5!(p6)$) {5};
    \node[point] (p13) at ($(p7)!0.5!(p5)$) {6};
    \node[point] (p14) at ($(p6)!0.5!(p7)$) {4};
    \path[topoedge0] ($(p5)$)--($(p11)$);
    \path[topoedge1] ($(p11)$)--($(p6)$);
    \path[topoedge0] ($(p6)$)--($(p14)$);
    \path[topoedge1] ($(p14)$)--($(p7)$);
    \path[topoedge0] ($(p7)$)--($(p13)$);
    \path[topoedge1] ($(p13)$)--($(p5)$);
    \node (start) [dof] at ($(p2)+\tikzmeshcelllen*.2*(.5,.5*\tikzsqrtthree)$) {};
    \draw[-latex, thick, black] (start) arc [orntarc, start angle=60, end angle=315];
    \tikzstyle{point}=[circle, minimum size=0pt,inner sep=0pt]
    \tikzstyle{point1}=[circle, minimum size=0pt,inner sep=0pt,fill=blue!40]
    \tikzstyle{orntarc}=[x radius=\tikzmeshcelllen*.2, y radius=\tikzmeshcelllen*.2]
    \tikzstyle{dof}=[circle,fill,inner sep=1.5pt]
    \tikzstyle{fsnode}=[circle,draw=black,minimum size=11pt,inner sep=0pt,fill=white]
    \tikzstyle{fsnode1}=[circle,draw=black,minimum size=11pt,inner sep=0pt,fill=blue!40]
    \tikzstyle{fsnode2}=[circle,draw=black,minimum size=11pt,inner sep=0pt,fill=red!40]
    \tikzstyle{topoedge0}=[draw=black,thick,shorten <= 0pt,shorten >= 0pt]
    \tikzstyle{topoedge1}=[draw=black,thick,shorten <= 0pt,shorten >= 0pt,-latex]
    \node (v0) [inner sep=0pt] at ($(0, 0)$) {};
    \node (v1) [inner sep=0pt] at ($\tikzmeshcelllen*(0, 1)$) {};
    \node (v2) [inner sep=0pt] at ($\tikzmeshcelllen*(1, 0)$) {};
    \node (v3) [inner sep=0pt] at ($\tikzmeshcelllen*(1, 1)$) {};
    \node (v4) [inner sep=0pt] at ($\tikzmeshcelllen*(1+\tikzsqrtthree/2, .5)$) {};
    \node[point] (p2) at ($\tikzmeshcelllen*(0, 0)+\tikzmeshcelllen*(1+.5/\tikzsqrtthree, .5)$) {};
    \node[point] (p5) at ($\tikzmeshcelllen*(0, 0)+(v2)$) {};
    \node[point] (p6) at ($\tikzmeshcelllen*(0, 0)+(v3)$) {};
    \node[point] (p7) at ($\tikzmeshcelllen*(0, 0)+(v4)$) {};
    \node[point] (p11) at ($(p5)!0.5!(p6)$) {};
    \node[point] (p13) at ($(p7)!0.5!(p5)$) {};
    \node[point] (p14) at ($(p6)!0.5!(p7)$) {};
    \path[topoedge0] ($(p5)$)--($(p6)$);
    \path[topoedge0] ($(p6)$)--($(p7)$);
    \path[topoedge0] ($(p7)$)--($(p5)$);
    \node[fsnode1] (fsnode3) at ($(p6)!0.25!(p7)$) {\footnotesize 3};
    \node[fsnode1] (fsnode4) at ($(p6)!0.50!(p7)$) {\footnotesize 4};
    \node[fsnode1] (fsnode5) at ($(p6)!0.75!(p7)$) {\footnotesize 5};
    \node[fsnode1] (fsnode6) at ($(p7)!0.25!(p5)$) {\footnotesize 6};
    \node[fsnode1] (fsnode7) at ($(p7)!0.50!(p5)$) {\footnotesize 7};
    \node[fsnode1] (fsnode8) at ($(p7)!0.75!(p5)$) {\footnotesize 8};
    \node[fsnode1] (fsnode10) at ($(p5)!0.25!(p6)$) {\footnotesize 10};
    \node[fsnode1] (fsnode11) at ($(p5)!0.50!(p6)$) {\footnotesize 11};
    \node[fsnode1] (fsnode12) at ($(p5)!0.75!(p6)$) {\footnotesize 12};
    \node[fsnode1] (fsnode0) at ($(fsnode10)!0.333!(fsnode5)$) {\footnotesize 0};
    \node[fsnode1] (fsnode1) at ($(fsnode10)!0.666!(fsnode5)$) {\footnotesize 1};
    \node[fsnode1] (fsnode2) at ($(fsnode11)!0.500!(fsnode4)$) {\footnotesize 2};
    \node[fsnode1] (fsnode9) at ($(p7)$) {\footnotesize 9};
    \node[fsnode1] (fsnode13) at ($(p6)$) {\footnotesize 13};
    \node[fsnode1] (fsnode14) at ($(p5)$) {\footnotesize 14};
    \node[point] (p2) at ($\tikzmeshcelllen*(2, 0)+\tikzmeshcelllen*(1+.5/\tikzsqrtthree, .5)$) {};
    \node[point] (p5) at ($\tikzmeshcelllen*(2, 0)+(v2)$) {};
    \node[point] (p6) at ($\tikzmeshcelllen*(2, 0)+(v3)$) {};
    \node[point] (p7) at ($\tikzmeshcelllen*(2, 0)+(v4)$) {};
    \node[point] (p11) at ($(p5)!0.5!(p6)$) {};
    \node[point] (p13) at ($(p7)!0.5!(p5)$) {};
    \node[point] (p14) at ($(p6)!0.5!(p7)$) {};
    \path[topoedge0] ($(p5)$)--($(p6)$);
    \path[topoedge0] ($(p6)$)--($(p7)$);
    \path[topoedge0] ($(p7)$)--($(p5)$);
    \node[fsnode1] (fsnode3) at ($(p6)!0.25!(p7)$) {\footnotesize 3};
    \node[fsnode1] (fsnode4) at ($(p6)!0.50!(p7)$) {\footnotesize 4};
    \node[fsnode1] (fsnode5) at ($(p6)!0.75!(p7)$) {\footnotesize 5};
    \node[fsnode2] (fsnode6) at ($(p7)!0.25!(p5)$) {\footnotesize 8};
    \node[fsnode1] (fsnode7) at ($(p7)!0.50!(p5)$) {\footnotesize 7};
    \node[fsnode2] (fsnode8) at ($(p7)!0.75!(p5)$) {\footnotesize 6};
    \node[fsnode1] (fsnode10) at ($(p5)!0.25!(p6)$) {\footnotesize 10};
    \node[fsnode1] (fsnode11) at ($(p5)!0.50!(p6)$) {\footnotesize 11};
    \node[fsnode1] (fsnode12) at ($(p5)!0.75!(p6)$) {\footnotesize 12};
    \node[fsnode2] (fsnode0) at ($(fsnode10)!0.333!(fsnode5)$) {\footnotesize 1};
    \node[fsnode2] (fsnode1) at ($(fsnode10)!0.666!(fsnode5)$) {\footnotesize 2};
    \node[fsnode2] (fsnode2) at ($(fsnode11)!0.500!(fsnode4)$) {\footnotesize 0};
    \node[fsnode1] (fsnode9) at ($(p7)$) {\footnotesize 9};
    \node[fsnode1] (fsnode13) at ($(p6)$) {\footnotesize 13};
    \node[fsnode1] (fsnode14) at ($(p5)$) {\footnotesize 14};
\end{tikzpicture}
\caption{
Example entity-local DoF orderings for P4 (CG4) finite element space.
(Left) An example physical mesh cell from \cref{Fi:function_dofs_saved} with the orderings of the cones indicated by the arrows.
(Middle) A schematic of DoFs correctly ordered on each entity relative to the cone of that entity.
(Right) A schematic of DoFs wrongly ordered;
DoFs on entities with local numbers 0 and 6 are ordered wrongly.
\label{Fi:orientation}
}
\end{figure}

\subsubsection{Local DoF vector on the mesh topology before saving}\label{SSS:saved_local_function}
Suppose that we have a finite element function space
on the mesh before saving, distributed over
$N$ parallel processes as in \cref{SS:mesh}.
Given the finite element,
we know the number of DoFs defined on the mesh topology entities before saving,
and denote by $\ol{D}_T^{n}$
the total number of DoFs defined on the entities visible to process $n\in\{0,\dots,N\minus 1\}$.
We regard that DoFs are \emph{owned} by process $n$
if they are defined on the entities owned by process $n$, and
call those that are not owned \emph{ghost} DoFs.
We denote by $\ol{\anarray{LocVEC}}_T^{n}$
the local vector on process $n$ containing
all the owned and the ghost DoF values;
such vectors containing all locally visible DoFs are
called \emph{local DoF vectors}.
We index this vector by
$\ol{j}_T^{n}\in\ol{\aset{J}}_T^{n}$ where
\begin{align}
\ol{\aset{J}}_T^{n}:=\{0,\dots,\ol{D}_T^{n}\minus 1\}.
\label{E:JTn}
\end{align}
In $\ol{\anarray{LocVEC}}_T^{n}$
we assume that DoFs defined on the same entity appear
in a contiguous chunk.
Firedrake by default chooses entity traversal order so that
parallel halo exchanges are handled easily, but
any entity traversal order can be used.

If multiple DoFs are defined on a given topological entity,
Firedrake orders them in $\ol{\anarray{LocVEC}}_T^{n}$
in a certain deterministic manner
relative to the ordering of the cone of that entity, or, in other words,
we can tell how those DoFs in the local DoF vector map
to members of the dual space
by inspecting the ordering of the cone of that entity.
In principle, we need to be able to determine such entity-local DoF ordering
on any given mesh topology entity for any given finite element.
Saving the finite element in addition to the discrete function space data
is necessary to retrieve such element-specific entity-local DoF orderings
as well as to recover the full representation of
the finite element function space when loading.

\Cref{Fi:function_saved} shows the example from \cref{Fi:mesh}
along with the function space DoF layout of
an example P4 (CG4) finite element space
using Firedrake's default entity traversal order and
entity-local DoF ordering.
It also details the corresponding $\ol{\anarray{LocVEC}}_T^{n}$ and $\ol{J}_T^{n}$,
$n\in\{0,1\}$.
\Cref{Fi:orientation} then gives more details on the entity-local DoF orderings
taking a mesh cell from \cref{Fi:function_saved} as an example.

\subsubsection{Local discrete function space data on the mesh topology before saving}\label{SSS:saved_local_function_space}
$\ol{\aset{I}}_T^{n}$ \cref{E:ITn} being the set of size $\ol{E}_T^{n}$ of local numbers
of entities visible to process $n\in\{0,\dots,N\minus 1\}$ of the mesh topology before saving, we define local integer arrays
$\ol{\anarray{LocG}}_T^{n}$,
$\ol{\anarray{LocDOF}}_T^{n}$, and $\ol{\anarray{LocOFF}}_T^{n}$,
all of size $\ol{E}_T^{n}$ and
indexed by $\ol{i}_T^{n}\in\ol{\aset{I}}_T^{n}$ \cref{E:ITn}, that contain
the corresponding global numbers,
number of DoFs defined on those entities, and
offsets in $\ol{\anarray{LocVEC}}_T^{n}$ to the first DoF values
on those entities, respectively.
We call discrete function space data that consist of these three integer arrays
\emph{local discrete function space data}.
Note that, to define the local discrete function space data, 
we do not need the local DoF vector;
we only need to know the finite element and the entity traversal order,
and thus we can define the local discrete function space data
before defining the local DoF vector.

\cref{Fi:function_data_saved} shows the
local discrete function space data,
associated with the example function space depicted
in \cref{Fi:function_dofs_saved}.

Note that, if DoFs are defined only on a subset of $\ol{I}_T^n$,
one can shrink the size of local discrete function space data by eliminating entities with no DoFs.
In practice, this will reduce the cost of parallel communication, and our
implementation in PETSc and Firedrake takes advantage of this.
However, in the following, as the nature of the algorithm does not change, we assume that DoFs are defined on all entities for simplicity of discussion.

\input{paper_fig_save_load_function}
\begin{figure}[ht]
\begin{tikzpicture}[darkstyle/.style={circle,draw,fill=gray!40}]
  \tikzstyle{cell0}=[]              
  \tikzstyle{cell1}=[fill=blue!40]  
  \tikzstyle{cell2}=[fill=orange!40]  
  \tikzstyle{setp}=[black, thick]
  \tikzstyle{subsetp0}=[black, thick, densely dotted]
  \tikzstyle{subsetp1}=[black, thick, densely dotted,fill=blue!40]
  \tikzstyle{subsetp2}=[black, thick, densely dotted,fill=orange!40]

  \node[inner sep=1pt] (XOO) at (0,0) {};
  \node[inner sep=1pt] (XO) at ($(XOO)+\tikzveccelllen*(7,0)$) {};
  \node[inner sep=1pt] () at ($(XO)+\tikzveccelllen*(-3,.5)$) {$\aset{L}_P$};
  \foreach \x in {0,1,...,14}
  {
    \ifthenelse{\x<5}
    {\node[inner sep=1pt] (XBL\x) at ($(XO)+\tikzveccelllen*(\x,0)$) {};} {}
    \ifthenelse{\x>4 \AND \x<10}
    {\node[inner sep=1pt] (XBL\x) at ($(XO)+\tikzveccelllen*(\x+1,0)$) {};} {}
    \ifthenelse{\x>9}
    {\node[inner sep=1pt] (XBL\x) at ($(XO)+\tikzveccelllen*(\x+2,0)$) {};} {}
    \node[inner sep=1pt] (XTR\x) at ($(XBL\x)+\tikzveccelllen*(1,1)$) {};
    \node (XT\x) at ($(XTR\x)-\tikzveccelllen*(.5,0)$) {};
    \node (XB\x) at ($(XBL\x)+\tikzveccelllen*(.5,0)$) {};
    \node (X\x) at ($(XBL\x)!0.5!(XTR\x)$) {};
  }
  \node[inner sep=1pt] (CO) at ($(XO)+\tikzveccelllen*(0, 5)$) {};
  \node[inner sep=1pt] (CLabel) at ($(CO)+\tikzveccelllen*(-3,.5)$) {$\aset{I}_P$};
  \foreach \x in {0,1,...,4}
  {
    \node[inner sep=1pt] (CBL\x) at ($(CO)+(\tikzveccelllen*\x, 0)$) {};
    \node[inner sep=1pt] (CTR\x) at ($(CO)+(\tikzveccelllen*\x+\tikzveccelllen, \tikzveccelllen)$) {};
    \node (C\x) at ($(CBL\x)!0.5!(CTR\x)$) {};
    \node (CT\x) at ($(CO)+(\tikzveccelllen*\x+\tikzveccelllen/2, \tikzveccelllen)$) {};
    \node (CB\x) at ($(CO)+(\tikzveccelllen*\x+\tikzveccelllen/2, 0)$) {};
  }
  \node[inner sep=1pt] (DO) at ($(CO)+\tikzveccelllen*(6,0)$) {};
  \foreach \x in {0,1,...,4}
  {
    \node[inner sep=1pt] (DBL\x) at ($(DO)+(\tikzveccelllen*\x, 0)$) {};
    \node[inner sep=1pt] (DTR\x) at ($(DO)+(\tikzveccelllen*\x+\tikzveccelllen, \tikzveccelllen)$) {};
    \node (D\x) at ($(DBL\x)!0.5!(DTR\x)$) {};
    \node (DT\x) at ($(DO)+(\tikzveccelllen*\x+\tikzveccelllen/2, \tikzveccelllen)$) {};
    \node (DB\x) at ($(DO)+(\tikzveccelllen*\x+\tikzveccelllen/2, 0)$) {};
    \node () at ($(D\x)$) {};
  }
  \node[inner sep=1pt] (EO) at ($(DO)+\tikzveccelllen*(6,0)$) {};
  \foreach \x in {0,1,...,4}
  {
    \node[inner sep=1pt] (EBL\x) at ($(EO)+(\tikzveccelllen*\x, 0)$) {};
    \node[inner sep=1pt] (ETR\x) at ($(EO)+(\tikzveccelllen*\x+\tikzveccelllen, \tikzveccelllen)$) {};
    \node (E\x) at ($(EBL\x)!0.5!(ETR\x)$) {};
    \node (ET\x) at ($(EO)+(\tikzveccelllen*\x+\tikzveccelllen/2, \tikzveccelllen)$) {};
    \node (EB\x) at ($(EO)+(\tikzveccelllen*\x+\tikzveccelllen/2, 0)$) {};
    \node () at ($(E\x)$) {};
  }
  \draw [subsetp0] ($(X2)$) ellipse [x radius=\tikzveccelllen*2.5, y radius=\tikzveccelllen*1];
  \draw [subsetp1] ($(X7)$) ellipse [x radius=\tikzveccelllen*2.5, y radius=\tikzveccelllen*1];
  \draw [subsetp2] ($(X12)$) ellipse [x radius=\tikzveccelllen*2.5, y radius=\tikzveccelllen*1];
  \draw [subsetp0] ($(C2)$) ellipse [x radius=\tikzveccelllen*2.5, y radius=\tikzveccelllen*1];
  \draw [subsetp1] ($(D2)$) ellipse [x radius=\tikzveccelllen*2.5, y radius=\tikzveccelllen*1];
  \draw [subsetp2] ($(E2)$) ellipse [x radius=\tikzveccelllen*2.5, y radius=\tikzveccelllen*1];
  \draw [setp] ($(X7)$) ellipse [x radius=\tikzveccelllen*9, y radius=\tikzveccelllen*1.75];
  \draw [setp] ($(D2)$) ellipse [x radius=\tikzveccelllen*9, y radius=\tikzveccelllen*1.75];
  \path[latex-, draw=black] (X1)--(C0);
  \path[latex-, draw=black] (X0)--(C1);
  \path[latex-, draw=black] (X4)--(C2);
  \path[latex-, draw=black] (X3)--(C3);
  \path[latex-, draw=black] (X12)--(C4);
  \path[latex-, draw=black] (X10)--(D0);
  \path[latex-, draw=black] (X9)--(D1);
  \path[latex-, draw=black] (X8)--(D2);
  \path[latex-, draw=black] (X2)--(D3);
  \path[latex-, draw=black] (X7)--(D4);
  \path[latex-, draw=black] (X6)--(E0);
  \path[latex-, draw=black] (X5)--(E1);
  \path[latex-, draw=black] (X14)--(E2);
  \path[latex-, draw=black] (X11)--(E3);
  \path[latex-, draw=black] (X13)--(E4);
  \foreach \x in {0,1,...,14}
  {
    \node () at ($(X\x)$) {\footnotesize \x};
  }
  \foreach \x in {0,1,...,4}
  {
    \node () at ($(C\x)$) {\footnotesize \x};
  }
  \foreach \x in {0,1,...,4}
  {
    \node () at ($(D\x)$) {\footnotesize \x};
  }
  \foreach \x in {0,1,...,4}
  {
    \node () at ($(E\x)$) {\footnotesize \x};
  }
  \node[anchor=west] () at ($(XOO)+\tikzveccelllen*(24,3)$) {$\chi_{\aset{I}_P}^{\aset{L}_P}$};
  \node[anchor=west] () at ($(C3)+\tikzveccelllen*(0,1)$) {$\aset{I}_P^{0}$};
  \node[anchor=west] () at ($(D3)+\tikzveccelllen*(0,1)$) {$\aset{I}_P^{1}$};
  \node[anchor=west] () at ($(E0)+\tikzveccelllen*(0,1)$) {$\aset{I}_P^{2}$};
  \node[anchor=west] () at ($(X2)+\tikzveccelllen*(-1,-2)$) {$\aset{L}_P^{0}$};
  \node[anchor=west] () at ($(X7)+\tikzveccelllen*(0,-2)$) {$\aset{L}_P^{2}$};
  \node[anchor=west] () at ($(X12)+\tikzveccelllen*(0,-2)$) {$\aset{L}_P^{1}$};
\end{tikzpicture}
\caption{
A schematic of the map $\chi_{\aset{I}_P}^{\aset{L}_P}$ from $\aset{I}_P$ to $\aset{L}_P$ associated with \cref{Fi:function_data_local_loaded}.
Parts owned by processes 0, 1, and 2 are shown in white, blue, and orange, respectively.
The inverse of this map is to be composed with $\chi_{\aset{I}_T}^{\aset{L}_P}$ depicted in \cref{Fi:mesh_map}.
\label{Fi:function_map}}
\end{figure}
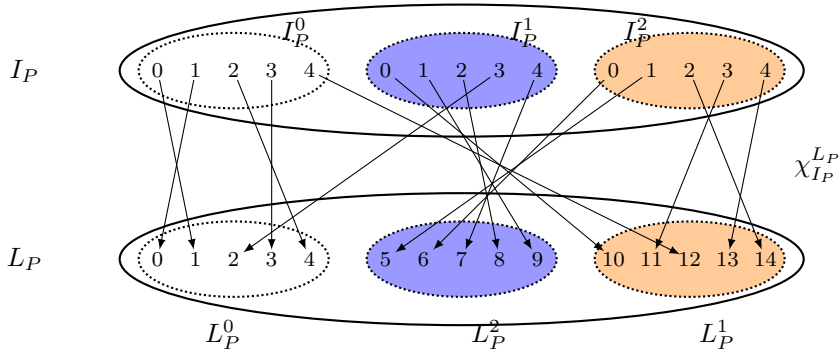

\subsubsection{Saving global DoF vectors}\label{SSS:saved_global_function}
We now consider saving the DoF vector.
We first convert the local DoF vector,
$\ol{\anarray{LocVEC}}_T^{n}$,
to the corresponding \emph{global DoF vector}
by excluding, on each process, all the ghost DoF values
and concatenating them across all processes.
The global DoF vector can be viewed as a serialised representation of the
DoF vector, but the vector is still partitioned.
In the actual implementation,
the concatenation is performed implicitly on disk, not in memory,
each process saving its own part of the global DoF vector concurrently to a file on a parallel file system.
We denote the global DoF vector by $\ol{\anarray{VEC}}_P$, whose indices, the global DoF indices, are collected in
\begin{align}
\aset{J}:=\{0,\dots,D\minus 1\},
\end{align}
where $D$ is the size of $\ol{\anarray{VEC}}_P$, or
the total number of DoFs.
Each process saving a DoF vector therefore amounts to saving the associated global DoF vector.

\Cref{Fi:function_data_global_saved} shows the global DoF vector $\ol{\anarray{VEC}}_P$
constructed from the local DoF vector $\ol{\anarray{LocVEC}}_T^{n}$, $n\in\{0,1\}$,
shown in \cref{Fi:function_data_saved}.

\subsubsection{Saving global discrete function space data}\label{SSS:saved_global_function_space}
Similarly, when we save the discrete function space data,
we first convert the local discrete function space data
to the corresponding \emph{global discrete function space data}
by removing, on each process, all entries associated with the ghost entities from 
$\ol{\anarray{LocG}}_T^{n}$, $\ol{\anarray{LocDOF}}_T^{n}$,
and $\ol{\anarray{LocOFF}}_T^{n}$.
The contents of $\ol{\anarray{LocOFF}}_T^{n}$ must further be adjusted
to contain the offsets in the global DoF vector $\ol{\anarray{VEC}}_P$
instead of those in the local DoF vector $\ol{\anarray{LocVEC}}_T^{n}$.
We then concatenate these arrays and save them as
the global discrete function space data.
We call these concatenated integer arrays
$\ol{\anarray{G}}_P$, $\ol{\anarray{DOF}}_P$, and $\ol{\anarray{OFF}}_P$,
and they are indexed by $\aset{I}$.

\cref{Fi:function_data_global_saved} shows
$\ol{\anarray{G}}_P$, $\ol{\anarray{DOF}}_P$, and $\ol{\anarray{OFF}}_P$
constructed from
$\ol{\anarray{LocG}}_T^{n}$, $\ol{\anarray{LocDOF}}_T^{n}$, and $\ol{\anarray{LocOFF}}_T^{n}$, $n\in\{0,1\}$,
shown in \cref{Fi:function_data_saved}.
Notice that a global offset of 20 has been added to $\ol{\anarray{LocOFF}}_T^{(1)}$ on concatenation
so that $\ol{\anarray{OFF}}_P$ contains offsets in the global vector $\ol{\anarray{VEC}}_P$.

\subsubsection{Loading global discrete function space data}\label{SSS:loaded_global_function_space}
We now consider loading the discrete function space data in parallel.
Recall that we denote by $N$ the number of saving processes, by $M$ the number of loading processes,
and, in general, $M\neq N$.
In this section we construct a map $\chi_{\aset{I}_P}^{\aset{L}_P}$ \cref{E:chi_IP_LP},
which associates the discrete function space data to the global numbers $\aset{I}$,
depicted in \cref{Fi:function_map} for our example.

The arrays representing the global discrete function space data
saved in \cref{SSS:saved_global_function_space} are first loaded in parallel onto $M$ processes.
For each array, chunks of nearly equal size (differing by one at most) are loaded on each process,
independently of how the array was saved from the original $N$ processes.
We denote these partitioned arrays by
$\anarray{G}_P$, $\anarray{DOF}_P$, and $\anarray{OFF}_P$.
These arrays are loaded onto the participating processes, and we denote parts of
$\anarray{G}_P$, $\anarray{DOF}_P$, and $\anarray{OFF}_P$
loaded onto process $m\in\{0,\dots,M\minus 1\}$ as
$\anarray{LocG}_P^{m}$, $\anarray{LocDOF}_P^{m}$, and $\anarray{LocOFF}_P^{m}$,
and index them all by $i_P^{m}\in\aset{I}_P^{m}$,
\begin{align}
\aset{I}_P^{m}:=\{0,\dots,E_P^{m}\minus 1\},
\label{E:IPm}
\end{align}
where $E_P^{m}$ is the size of
$\anarray{LocG}_P^{m}$ (equal to size of $\anarray{LocDOF}_P^{m}$ and $\anarray{LocOFF}_P^{m}$) and
$\sum_{m\in\{0,\dots,M\minus 1\}}E_P^m=E$.
For convenience, we define a union set $\aset{I}_P$ as
\begin{align}
\aset{I}_P:=
\bigcup\limits_{m\in\{0,\dots,M\minus 1\}}\{m\}\times\aset{I}_P^{m}.
\label{E:IP}
\end{align}
Once $\anarray{LocG}_P^{m}$, an array of global numbers in $\aset{I}$
indexed by $\aset{I}_P^{m}$,
has been loaded to each process $m$,
one can use the partition formula \cref{E:chi_E_LP}
to construct a bijective map
from $\aset{I}_P$ \cref{E:IP} to $\aset{L}_P$ \cref{E:LP},
\begin{align}
\chi_{\aset{I}_P}^{\aset{L}_P}:\aset{I}_P\rightarrow\aset{L}_P.
\label{E:chi_IP_LP}
\end{align}
We are specifically interested in
the inverse of $\chi_{\aset{I}_P}^{\aset{L}_P}$ \cref{E:chi_IP_LP}, and
we are to represent this inverse as a star forest
with $\aset{I}_P$ and $\aset{L}_P$ representing roots and leaves.

\cref{Fi:function_data_global_loaded} shows example
$\anarray{G}_P$, $\anarray{DOF}_P$, and $\anarray{OFF}_P$
partitioned for three parallel processes.
Notice the arrays are all partitioned in the same fashion.
\Cref{Fi:function_data_local_loaded} shows the local views of the
partitioned arrays,
$\anarray{LocG}_P^{m}$, $\anarray{LocDOF}_P^{m}$, and $\anarray{LocOFF}_P^{m}$,
along with the local indices $\aset{I}_P^{m}$, for $m\in\{0,1,2\}$.
\Cref{Fi:function_map} shows a schematic of the associated map
$\chi_{\aset{I}_P}^{\aset{L}_P}$.

\subsubsection{Loading global DoF vectors}\label{SSS:loaded_global_function}
When we are loading the global DoF vector saved in \cref{SSS:saved_global_function},
it is again partitioned completely independently from the partitioning used when saving.
We denote the resulting partitioned global DoF vector by $\anarray{VEC}_P$.
Part of $\anarray{VEC}_P$ loaded onto process $m\in\{0,\dots,M\minus 1\}$
is denoted by $\anarray{LocVEC}_P^{m}$ and
locally indexed by $j_P^{m}\in\aset{J}_P^{m}$,
\begin{align}
\aset{J}_P^{m}:=\{0,\dots,D_P^{m}\minus 1\},
\label{E:JPm}
\end{align}
where $D_P^{m}$ is the size of $\anarray{LocVEC}_P^{m}$.
Thus, $\sum_{m\in\{0,\dots,M\minus 1\}}D_P^{m}=D$, the total number of DoFs.
For convenience, we define a union set
\begin{align}
\aset{J}_P:=
\bigcup\limits_{m\in\{0,\dots,M\minus 1\}}\{m\}\times\aset{J}_P^{m}.
\label{E:JP}
\end{align}
We denote the simple bijective map from the set of global DoF indices,
$\{0,\dots,\aset{D}\minus 1\}$,
to $\aset{J}_P$ \cref{E:JP} as
\begin{align}
\chi_{\{0,\dots,\aset{D}-1\}}^{\aset{J}_P}:
\{0,\dots,\aset{D}\minus 1\}\rightarrow\aset{J}_P,
\label{E:chi_D_JP}
\end{align}
which represents the partition of the global DoF vector.

\cref{Fi:function_data_global_loaded} also shows example $\anarray{VEC}_P$
partitioned for three parallel processes.
\cref{Fi:function_data_local_loaded} shows the local views
$\anarray{LocVEC}_P^{m}$
along with the sets of local indices $\aset{J}_P^{m}$
for $m\in\{0,1,2\}$.

\subsubsection{Notes}\label{SSS:notes}
In practice one only needs to save
the discrete function space data
$\ol{\anarray{G}}_P$, $\ol{\anarray{DOF}}_P$, $\ol{\anarray{OFF}}_P$, and the
representation of the finite element once.
One can then save as many DoF vectors associated with the discrete function space data,
represented by $\ol{\anarray{VEC}}_P$,
as needed without repeatedly saving the same discrete function space data,
which is often desirable, particularly for time-dependent problems.
The same holds for loading; all those data need to be loaded only once.

\subsection{Reconstructing function spaces and functions on the loaded mesh}\label{SS:reconstruct}
We described how we load the mesh topology data and
reconstruct the mesh topology in \cref{SS:mesh}, and
how we load the discrete function space data
and associated DoF vectors
independently of the topology data in \cref{SS:function}.
In this section, we broadcast the loaded
discrete function space data and associated DoF vectors
onto the loaded mesh topology
to reconstruct the function space and associated functions
on the loaded topology,
making use of the sets and maps constructed in
\cref{SSS:loaded_mesh},
\cref{SSS:loaded_global_function_space}, and
\cref{SSS:loaded_global_function}.

\subsubsection{Local discrete function space data on the loaded mesh topology}\label{SSS:loaded_local_function_space}
In this section, we use the maps
$\chi_{\aset{I}_T}^{\aset{L}_P}$ \cref{E:chi_IT_LP},
associating entities on the loaded mesh topology to the global numbers $\aset{I}$, and
$\chi_{\aset{I}_P}^{\aset{L}_P}$ \cref{E:chi_IP_LP},
associating the discrete function space data to the same set of global numbers,
to reconstruct the local discrete function space data on the loaded parallel mesh topology.
The local discrete function space data are composed of
$\anarray{LocG}_T^{m}$, $\anarray{LocDOF}_T^{m}$, and $\anarray{LocOFF}_T^{m}$,
all indexed by $i_T^{m}\in\aset{I}_T^{m}$ \cref{E:ITm}
for $m\in\{0,\dots,M\minus 1\}$.
These arrays all have a size of $E_T^{m}$.
As $\anarray{LocG}_T^{m}$ was constructed when we loaded the mesh topology in \cref{SSS:loaded_mesh},
we only discuss construction of $\anarray{LocDOF}_T^{m}$ and $\anarray{LocOFF}_T^{m}$.
For convenience, we define $\anarray{DOF}_T$ and $\anarray{OFF}_T$ as collections of
$\anarray{LocDOF}_T^m$ and $\anarray{LocOFF}_T^m$, respectively, $m\in\{0,\dots,M\minus 1\}$,
and introduce shorthand notations
\begin{align}
\anarray{ARRAY}_X[i_X]&:=\anarray{LocARRAY}_X^{m}[i_X^{m}],
\end{align}
where $X$ stands for $T$ or $P$,
$i_X=(m,i_X^{m})\in\aset{I}_X$,
and
$\anarray{ARRAY}$ stands for
$\anarray{G}$, $\anarray{DOF}$, or $\anarray{OFF}$.

We first focus on $\anarray{DOF}_T$, indexed by $\aset{I}_T$ \cref{E:IT}.
Number of DoFs defined on each entity
was loaded into $\anarray{DOF}_P$, indexed by $\aset{I}_P$ \cref{E:IP}, in \cref{SSS:loaded_global_function_space}.
Thus, to broadcast $\anarray{DOF}_P$ to $\anarray{DOF}_T$,
one requires a map $\chi_{\aset{I}_T}^{\aset{I}_P}$
from $\aset{I}_T$ to $\aset{I}_P$,
which can be readily constructed using
the bijective map
$\chi_{\aset{I}_P}^{\aset{L}_P}$ \cref{E:chi_IP_LP} and
the surjective map
$\chi_{\aset{I}_{T}}^{{\aset{L}_P}}$ \cref{E:chi_IT_LP}
as
\begin{align}
\chi_{\aset{I}_T}^{\aset{I}_P}:=
(\chi_{\aset{I}_P}^{\aset{L}_P})^{-1}\circ
\chi_{\aset{I}_{T}}^{{\aset{L}_P}}.
\label{E:chi_IT_IP}
\end{align}
We can then use this map to do the broadcast,
\begin{align}
\anarray{DOF}_T[i_T]&=
\anarray{DOF}_P[\chi_{\aset{I}_T}^{\aset{I}_P}(i_T)],
\label{E:DOF_bcast}
\end{align}
for each $i_T=(m,i_T^{m})\in\aset{I}_T$.

Now, let us deal with $\anarray{OFF}_T$.
As $\anarray{LocDOF}_T^{m}[i_T^{m}]$ is
the number of DoFs defined on the entity with local number $i_T^{m}$,
we now know the size $D_T^{m}$ of the local DoF vector $\anarray{LocVEC}_T^{m}$:
$D_T^{m}=\sum_{i_T^{m}\in\aset{I}_T^{m}}\anarray{LocDOF}_T^{m}[i_T^{m}]$.
We index ${\anarray{LocVEC}_T^{m}}$
with $j_T^{m}\in\aset{J}_T^{m}$,
\begin{align}
\aset{J}_T^{m}:=\{0,\dots,D_T^{m}\minus 1\},
\label{E:JTm}
\end{align}
and define a union set
\begin{align}
\aset{J}_{T}:=
\bigcup\limits_{m\in\{0,\dots,M\minus 1\}}\{m\}\times\aset{J}_T^{m}.
\label{E:JT}
\end{align}
Similarly to the local DoF vectors before saving,
DoF values in $\anarray{LocVEC}_T^{m}$ are ordered per following rules,
which determine the $\anarray{LocOFF}_T^{m}$ array:
\begin{enumerate}
    \item DoFs defined on the same entity appear in a contiguous chunk
in $\anarray{LocVEC}_T^{m}$,
    \item DoFs are collected by traversing all entities represented by local numbers in
${\aset{I}_T^{m}}$ in some order, and,
    \item if multiple DoFs are defined on an entity,
the order in which they appear in the contiguous chunk of
$\anarray{LocVEC}_T^{m}$
is determined by the cone of that entity
as explained in \cref{SS:function}.
\end{enumerate}
Note that, in general, the rule for the entity traversal order is completely arbitrary
on the loaded topology and can be different from the one used for the saved local DoF vectors.
Firedrake, however, by default, takes steps to ensure that the order is the same if one reloads on the same number of processors.
Note also that the order of the lower-dimensional entities appearing
in the cone of each entity is preserved in the save-load cycle,
so, if DoFs on an entity are laid out in reference to
the cone of that entity,
there will be no ambiguity in determining
which DoF value in $\anarray{LocVEC}_T^{m}$
is associated with which member of the dual space.

\input{paper_fig_loaded_function}

\Cref{Fi:loaded} depicts the loaded mesh topology
first shown in \cref{Fi:mesh_loaded},
the P4 (CG4) function space from \cref{Fi:function_dofs_saved}
migrated onto the loaded mesh,
and arrays
representing the local discrete function space data
$\anarray{LocG}_T^{m}$, $\anarray{LocDOF}_T^{m}$, and $\anarray{LocOFF}_T^{m}$ for $m\in\{0,1,2\}$.

\subsubsection{Local DoF vector on the loaded mesh topology}\label{SSS:loaded_local_function}
We are now ready to fill in the local DoF vector.
As in \cref{SSS:loaded_local_function_space},
we define $\anarray{VEC}_T$ as a collection of $\anarray{LocVEC}_T^m$, $m\in\{0,\dots,M\minus 1\}$,
and introduce shorthand notations
\begin{align}
\anarray{VEC}_X[j_X]:=\anarray{LocVEC}_X^{m}[j_X^{m}],
\end{align}
where $X$ stands for $T$ or $P$ and $j_X=(m,j_X^{m})\in\aset{J}_X$.

Our goal in this section is to fill in $\anarray{VEC}_T$,
which is indexed by $\aset{J}_T$ \cref{E:JT}.
The saved DoF vector data were loaded onto $\anarray{VEC}_P$ in \cref{SSS:loaded_global_function},
which is indexed by $\aset{J}_P$ \cref{E:JP},
so we need to construct a map $\chi_{\aset{J}_T}^{\aset{J}_P}$
from $\aset{J}_T$ to $\aset{J}_P$.
We first construct a surjective map
$\chi_{\aset{J}_T}^{\aset{J}}:\aset{J}_T\rightarrow\aset{J}$.
For each $j_T\in\aset{J}_T$, there exists $m\in\{0,\dots,M\minus 1\}$ and $j_T^{m}\in\aset{J}_T^{m}$
such that $j_T=(m,j_T^{m})\in\aset{J}_T$.
Given $j_T^m$, one can find
$i_T^{m}\in\aset{I}_T^{m}$ and $i_\textrm{DoF}\in\{0,\dots,\anarray{LocDOF}_T^{m}[i_T^{m}]\minus 1\}$
such that 
$j_T^{m}=\anarray{LocOFF}_T^{m}[i_T^{m}]+i_\textrm{DoF}$.
Then one can construct the surjective map
\begin{align}
\chi_{\aset{J}_T}^{\aset{J}}(j_T)=
\anarray{OFF}_P[\chi_{\aset{I}_T}^{\aset{I}_P}((m,i_T^{m}))]+i_\textrm{DoF},
\label{E:chi_JT_D_jT}
\end{align}
where $\anarray{OFF}_P$ was loaded in \cref{SSS:loaded_global_function_space} and $\chi_{\aset{I}_T}^{\aset{I}_P}$ is defined in ~\cref{E:chi_IT_IP}.
We then compose
the bijective partition map $\chi_{\aset{J}}^{\aset{J}_P}$ \cref{E:chi_D_JP} with
this surjective map $\chi_{\aset{J}_T}^{\aset{J}}$ \cref{E:chi_JT_D_jT} to obtain
\begin{align}
\chi_{\aset{J}_T}^{\aset{J}_P}&:=
\chi_{\aset{J}}^{\aset{J}_P}\circ
\chi_{\aset{J}_T}^{\aset{J}}.
\label{E:chi_JT_JP}
\end{align}
Finally, using $\chi_{\aset{J}_T}^{\aset{J}_P}$ \cref{E:chi_JT_JP}, the broadcasting formula for the local DoF vector is given by
\begin{align}
\anarray{VEC}_T[j_T]=
\anarray{VEC}_P[\chi_{\aset{J}_T}^{\aset{J}_P}(j_T)].
\label{E:VEC_bcast}
\end{align}

\Cref{Fi:function_data_loaded} shows the local DoF vector $\anarray{LocVEC}_T^{m}$ attached to the loaded mesh topology for $m\in\{0,1,2\}$.
Note that the DoF values were broadcast from the $\anarray{VEC}_P$ array depicted in \cref{Fi:function_data_global_loaded}.

\section{Implementation}\label{S:impl}
In this section, we describe our implementation of the algorithm
outlined in \cref{S:concept} in Firedrake~\cite{FiredrakeUserManual, Rathgeber2016} and PETSc~\cite{petsc-user-ref, petsc-web-page}.
Firedrake is a high-level, high-productivity system for the specification and solution of partial differential equations (PDEs) using the finite element method.
The core data structures used to represent solutions of PDEs, such as the velocity and pressure in a fluid simulation, are the combination of a mesh of the domain of interest, a finite element function space, and a function belonging to the discrete function space.
Firedrake uses PETSc to represent these data structures;
Firedrake meshes, function spaces, and functions are almost directly represented by PETSc \objname{DMPlex}, \objname{PetscSection}, and \objname{Vec} objects, respectively.
Therefore, we first write functions in PETSc for saving and loading these objects and then wrap these in a new \objname{CheckpointFile} class in Firedrake along with Firedrake-specific metadata.

\subsection{Meshes}\label{SS:mesh_impl}
Our algorithm for mesh saving and loading has been introduced in \cref{SS:mesh}.
Firedrake uses parallel mesh data structures called
\objname{DMPlex}~\cite{KnepleyKarpeev09,LangeMitchellKnepleyGorman2015,LangeKnepleyGorman2015}
provided by PETSc to describe mesh topologies.
A \objname{DMPlex} represents entities such as vertices, edges, faces, and cells, and their adjacency,
which can be described by cone relations introduced in \cref{SS:mesh}.
The cone relations form edges of a \emph{directed acyclic graph} (DAG),
referred to in mathematics as a Hasse diagram~\cite[Ch.~1
\S 5]{birkhoff1948}.
Mesh entities are represented by vertices in this graph and called \emph{DAG points} in this context.

In a local \objname{DMPlex} mesh,
each mesh entity carries a distinct local number
drawn from, e.g., $\aset{I}_T^{m}$.
A parallel mesh is simply a collection of \objname{DMPlex} objects
combined with a map relating points of the mesh on one process to that on another,
such as shared vertices along a process boundary.
This map is encoded in an instance of
\objname{PetscSF}~\cite{PetscSF2022},
a PETSc implementation of the star forest concept introduced in \cref{S:concept}.
A \objname{PetscSF}  describes a parallel map from \emph{leaves} to \emph{roots};
a root is a local number (DoF index) on some process and
an associated leaf is a local number (DoF index) on the same or another process.
Multiple leaves can be associated with the same root,
and this relation means they would represent the same entity in a monolithic mesh.
The \objname{PetscSF} stored in the \objname{DMPlex} object mentioned above
is called \objname{pointSF}.
We also employ the \objname{PetscSFCompose} function, which composes
two \objname{PetscSF} objects, creating a new \objname{PetscSF}, and
\objname{PetscSFBcast} to broadcast data associated with roots
to the corresponding leaves.

Each parallel mesh entity is assigned a global number
in $\aset{I}$, where $E$ is the total number of
unique topological points on the entire mesh,
using the \objname{DMPlexCreatePointNumbering()} function;
one can then associate each local number in the local mesh with a global number.
\begin{figure}[ht]
\centering
\begin{tikzpicture}
    \tikzstyle{neuron0}=[circle, draw=black, minimum size=13pt,inner sep=0pt]
    \tikzstyle{neuron1}=[circle, draw=black, minimum size=13pt,inner sep=0pt,fill=blue!40]
    \tikzstyle{neuron2}=[circle, draw=black, minimum size=13pt,inner sep=0pt,fill=orange!40]
    \tikzstyle{dagedge}=[draw=black,latex-,shorten <= 0pt,shorten >= 0pt]
    \node[] (O) at (0, 0) {};
    \node[neuron0] (p0) at ($(O)+(0*\tikzoffx, 0)$) {$0$};
    \node[neuron1] (p1) at ($(p0)+(\tikzoffx, 0)$) {$1$};
    \node[neuron0] (p2) at ($(p0)+(-\tikzoffx, 2*\tikzoffy)$) {$2$};
    \node[neuron0] (p3) at ($(p0)+(0*\tikzoffx, 2*\tikzoffy)$) {$3$};
    \node[neuron0] (p4) at ($(p0)+(1*\tikzoffx, 2*\tikzoffy)$) {$4$};
    \node[neuron1] (p5) at ($(p0)+(2*\tikzoffx, 2*\tikzoffy)$) {$5$};
    \node[neuron0] (p6) at ($(p0)+(-1.5*\tikzoffx, \tikzoffy)$) {$6$};
    \node[neuron0] (p7) at ($(p0)+(-0.5*\tikzoffx, \tikzoffy)$) {$7$};
    \node[neuron0] (p8) at ($(p0)+( 0.5*\tikzoffx, \tikzoffy)$) {$8$};
    \node[neuron1] (p9) at ($(p0)+( 1.5*\tikzoffx, \tikzoffy)$) {$9$};
    \node[neuron1] (p10) at ($(p0)+(2.5*\tikzoffx, \tikzoffy)$) {$10$};

    \path[dagedge] (p0)--(p6);
    \path[dagedge] (p0)--(p7);
    \path[dagedge] (p0)--(p8);
    \path[dagedge] (p1)--(p7);
    \path[dagedge] (p1)--(p9);
    \path[dagedge] (p1)--(p10);
    \path[dagedge] (p6)--(p2);
    \path[dagedge] (p6)--(p3);
    \path[dagedge] (p7)--(p4);
    \path[dagedge] (p7)--(p3);
    \path[dagedge] (p8)--(p2);
    \path[dagedge] (p8)--(p4);
    \path[dagedge] (p9)--(p3);
    \path[dagedge] (p9)--(p5);
    \path[dagedge] (p10)--(p4);
    \path[dagedge] (p10)--(p5);
    \node[circle, minimum size=18pt] (p0) at ($(O)+(6.5*\tikzoffx, 0)$) {};
    \node[circle, minimum size=18pt] (p2) at ($(p0)+(-2*\tikzoffx, \tikzoffy)$) {};
    \node[circle, minimum size=18pt] (p3) at ($(p0)+(0*\tikzoffx, \tikzoffy)$) {};
    \path[rounded corners=1mm,fill=gray!20,inner sep=10pt] (p0.south)--(p0.south west)--(p2.south west)--(p2.west)--(p2.north west)--(p2.north)--(p3.north)--(p3.north east)--(p3.east)--(p0.east)--(p0.south east)--cycle {};
    \node[neuron2] (p0) at ($(O)+(6.5*\tikzoffx, 0)$) {$0$};
    \node[neuron1] (p1) at ($(p0)+(\tikzoffx, 0)$) {$1$};
    \node[neuron0] (p2) at ($(p1)+(\tikzoffx, 0)$) {$2$};
    \node[neuron2] (p3) at ($(p0)+(-\tikzoffx, 2*\tikzoffy)$) {$3$};
    \node[neuron1] (p4) at ($(p3)+(\tikzoffx,0)$) {$4$};
    \node[neuron0] (p5) at ($(p4)+(\tikzoffx,0)$) {$5$};
    \node[neuron0] (p6) at ($(p5)+(\tikzoffx,0)$) {$6$};
    \node[neuron0] (p7) at ($(p6)+(\tikzoffx,0)$) {$7$};
    \node[neuron2] (p8) at ($(p0)+(-2*\tikzoffx,\tikzoffy)$) {$8$};
    \node[neuron1] (p9) at ($(p8)  +(\tikzoffx,0)$) {$9$};
    \node[neuron2] (p10) at ($(p9) +(\tikzoffx,0)$) {$10$};
    \node[neuron0] (p11) at ($(p10)+(\tikzoffx,0)$) {$11$};
    \node[neuron1] (p12) at ($(p11)+(\tikzoffx,0)$) {$12$};
    \node[neuron0] (p13) at ($(p12)+(\tikzoffx,0)$) {$13$};
    \node[neuron0] (p14) at ($(p13)+(\tikzoffx,0)$) {$14$};

    \path[dagedge] (p0)--(p8);
    \path[dagedge] (p0)--(p9);
    \path[dagedge] (p0)--(p10);
    \path[dagedge] (p1)--(p11);
    \path[dagedge] (p1)--(p9);
    \path[dagedge] (p1)--(p12);
    \path[dagedge] (p2)--(p13);
    \path[dagedge] (p2)--(p11);
    \path[dagedge] (p2)--(p14);
    \path[dagedge] (p8)--(p4);
    \path[dagedge] (p8)--(p3);
    \path[dagedge] (p9)--(p5);
    \path[dagedge] (p9)--(p4);
    \path[dagedge] (p10)--(p3);
    \path[dagedge] (p10)--(p5);
    \path[dagedge] (p11)--(p6);
    \path[dagedge] (p11)--(p5);
    \path[dagedge] (p12)--(p6);
    \path[dagedge] (p12)--(p4);
    \path[dagedge] (p13)--(p7);
    \path[dagedge] (p13)--(p5);
    \path[dagedge] (p14)--(p7);
    \path[dagedge] (p14)--(p6);
    \node[neuron1] (p0) at ($(O)+(14*\tikzoffx, 0)$) {$0$};
    \node[neuron2] (p1) at ($(p0)+(\tikzoffx, 0)$) {$1$};
    \node[neuron0] (p2) at ($(p0)+(-\tikzoffx, 2*\tikzoffy)$) {$2$};
    \node[neuron0] (p3) at ($(p0)+(0*\tikzoffx, 2*\tikzoffy)$) {$3$};
    \node[neuron1] (p4) at ($(p0)+(1*\tikzoffx, 2*\tikzoffy)$) {$4$};
    \node[neuron2] (p5) at ($(p0)+(2*\tikzoffx, 2*\tikzoffy)$) {$5$};
    \node[neuron0] (p6) at ($(p0)+(-1.5*\tikzoffx, \tikzoffy)$) {$6$};
    \node[neuron1] (p7) at ($(p0)+(-0.5*\tikzoffx, \tikzoffy)$) {$7$};
    \node[neuron1] (p8) at ($(p0)+( 0.5*\tikzoffx, \tikzoffy)$) {$8$};
    \node[neuron2] (p9) at ($(p0)+( 1.5*\tikzoffx, \tikzoffy)$) {$9$};
    \node[neuron2] (p10) at ($(p0)+(2.5*\tikzoffx, \tikzoffy)$) {$10$};

    \path[dagedge] (p0)--(p6);
    \path[dagedge] (p0)--(p7);
    \path[dagedge] (p0)--(p8);
    \path[dagedge] (p1)--(p9);
    \path[dagedge] (p1)--(p7);
    \path[dagedge] (p1)--(p10);
    \path[dagedge] (p6)--(p2);
    \path[dagedge] (p6)--(p3);
    \path[dagedge] (p7)--(p3);
    \path[dagedge] (p7)--(p4);
    \path[dagedge] (p8)--(p2);
    \path[dagedge] (p8)--(p4);
    \path[dagedge] (p9)--(p4);
    \path[dagedge] (p9)--(p5);
    \path[dagedge] (p10)--(p5);
    \path[dagedge] (p10)--(p3);
\end{tikzpicture}
\caption{DAG representation of the parallel mesh topology
shown in \cref{Fi:mesh_loaded} (repeated in \cref{Fi:mesh_loaded_copy})
that one can represent with a parallel \objname{DMPlex}.
Local numbers are shown for local topologies on processes 0, 1, and 2, from left to right.
Entities owned by processes 0, 1, and 2 are shown in white, blue, and orange.
The cell with local number 0 and its cone on process 1 are highlighted.
\label{Fi:dag}}
\end{figure}
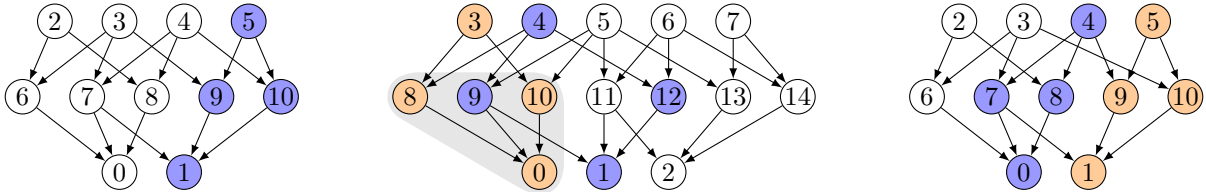
\cref{Fi:dag} shows the parallel \objname{DMPlex} object
that represents the loaded mesh topology shown in
\cref{Fi:mesh_loaded} and repeated in \cref{Fi:mesh_loaded_copy}
as an example.
Using this flexible representation,
\objname{DMPlex} can manipulate meshes of any dimension,
any combination of cell shapes,
and even geometrically non-conforming meshes
with hanging nodes~\cite{IsaacKnepley2017}.
Although not directly used in this work,
\objname{DMPlex} decorates each edge of the DAG with an orientation number.
This identifies the member of the dihedral group for each subcell (face)
used to transform it from the canonical ordering before attaching it to a given cell.

As discussed in \cref{SS:mesh}, when saving a mesh topology,
we save cones written in terms of global numbers.
This is done by \objname{DMPlexTopologyView} function.
\objname{DMPlexTopologyView} is wrapped in the \objname{save\_mesh} method
of \objname{CheckpointFile} class in Firedrake.

Conversely, we use \objname{DMPlexTopologyLoad} function
to load and reconstruct the saved mesh topology.
Once the DAG has been reconstructed and the \objname{pointSF} is created locally
on each process,
local numbers representing $\aset{I}_T^{m}$,
and thus $\aset{I}_T$ \cref{E:IT}, can be identified.
As cones written in terms of global numbers were used
to reconstruct the DAG,
one can readily construct an array of
global numbers representing $\anarray{LocG}_T^{m}$.
Knowing the total number of unique topological points, $E$,
we choose the partition formula \cref{E:chi_E_LP},
construct $\aset{L}_P^{m}$, and then $\aset{L}_P$ \cref{E:LP}.
After redistributing the loaded \objname{DMPlex} and adding partition overlaps if needed
as described in \cref{S:loaded_mesh_details},
we have a \objname{PetscSF} representing $\chi_{\aset{I}_T}^{\aset{L}_P}$ \cref{E:chi_IT_LP}.
\objname{DMPlexTopologyLoad}
is wrapped in the \objname{load\_mesh} method of \objname{CheckpointFile} class.

Finally, in checkpoint-and-restart applications, one might desire the loaded mesh to have the exact same parallel distribution as the mesh before saving.
Thus, in PETSc, we included a feature to allow for recovering the exact same parallel distribution when the loading process count is the same as the saving process count, and
let Firedrake use that feature by default.
If this feature is enabled, the saved arrays representing the mesh distribution are loaded and directly used to distribute the mesh across processes upon loading.

\subsection{Function spaces}\label{SS:function_space_impl}
We now describe how Firedrake function spaces are saved and loaded
following the discussion in \cref{SS:function} and \cref{SS:reconstruct}.
Firedrake describes a finite element function space using a mesh
and a \emph{UFL element},
a symbolic representation of a finite element provided by
the Unified Form Language (UFL)~\cite{ufl2014}, and thus
Firedrake function space is represented by a UFL element and
a representation of the discrete function space data.
As saving and loading a serialised UFL element is straightforward, we focus on saving and loading the discrete function space data.

Local and global discrete function space data are
exactly represented by
\objname{PetscSection} objects
defined on the \objname{DMPlex} mesh topology,
called the \emph{local} section and the \emph{global} section, respectively.
Namely, the local section along with the underlying \objname{DMPlex} topology can represent
$\ol{\anarray{LocG}}_T^{m}$,
$\ol{\anarray{LocDOF}}_T^{m}$, and
$\ol{\anarray{LocOFF}}_T^{m}$, and
the global section along with the underlying \objname{DMPlex} topology can represent
$\ol{\anarray{G}}_P$,
$\ol{\anarray{DOF}}_P$, and
$\ol{\anarray{OFF}}_P$.
Note one can construct the global section from the local section using \objname{pointSF}.

The \objname{DMPlexSectionView} function is used to save the section data.
It first constructs the global section from the local section,
if it has not been done yet;
specifically, it makes a global section representing
$\ol{\anarray{G}}_P$,
$\ol{\anarray{DOF}}_P$, and
$\ol{\anarray{OFF}}_P$ from the local section representing
$\ol{\anarray{LocG}}_T^{m}$,
$\ol{\anarray{LocDOF}}_T^{m}$, and
$\ol{\anarray{LocOFF}}_T^{m}$.
\objname{DMPlexSectionView} function then saves arrays representing
the global section
as the representation of the global discrete function space data
as illustrated in \cref{SS:function}.

The first time the \objname{save\_function} method of Firedrake's \objname{CheckpointFile} class is called for a function in a particular function space, 
it calls \objname{DMPlexSectionView} and also saves the UFL element (see \cref{SS:function_impl}).

Conversely, the \objname{DMPlexSectionLoad} function
reconstructs the local and global sections
on the loaded \objname{DMPlex} topology by first constructing a \objname{PetscSF} representing
$\chi_{\aset{I}_T}^{\aset{I}_P}$ \cref{E:chi_IT_IP}.
It then constructs a \objname{PetscSF} representing
$\chi_{\aset{J}_T}^{\aset{J}_P}$ \cref{E:chi_JT_JP}
that can be used to load functions as described in \cref{SS:function_impl}.

Following the procedure outlined in \cref{SS:function}
we first load the global section data into the arrays representing
$\anarray{LocG}_P^{m}$, $\anarray{LocDOF}_P^{m}$, and $\anarray{LocOFF}_P^{m}$,
and identifies $\aset{I}_P^{m}$, which indexes these arrays, and then
$\aset{I}_P$ \cref{E:IP}.
Using the array representing $\anarray{LocG}_P^{m}$
along with the map
$\chi_{\aset{I}}^{\aset{L}_P}$ \cref{E:chi_E_LP} that were chosen
when we loaded the topology,
we can construct a \objname{PetscSF} representing the \emph{inverse} of
$\chi_{\aset{I}_P}^{\aset{L}_P}$ \cref{E:chi_IP_LP}.
Using the \objname{PetscSF} representing $\chi_{\aset{I}_T}^{\aset{L}_P}$ \cref{E:chi_IT_LP},
constructed when we loaded the topology,
a \objname{PetscSF} representing
$\chi_{\aset{I}_T}^{\aset{I}_P}$ \cref{E:chi_IT_IP}
can be constructed using the \objname{PetscSFCompose} function.
An array representing
$\anarray{LocDOF}_T^{m}$ is then constructed
using the broadcasting formula \cref{E:DOF_bcast}
using the \objname{PetscSFBcast} function.
An array representing
$\anarray{LocOFF}_T^{m}$ can then be constructed independently
as mentioned in \cref{SS:reconstruct}.
With the above, we have now reconstructed the local section on the
loaded \objname{DMPlex} topology;
the global section can also be constructed from the local section if necessary.

When we load the function space, we also construct $\chi_{\aset{J}_T}^{\aset{J}_P}$ \cref{E:chi_JT_JP},
which is to be used when loading functions defined on the function space.
Computing the number of DoFs in the local DoF vector, $D_T^{m}$,
using the local section, we identify
$\aset{J}_T^{m}$, and then $\aset{J}_T$ \cref{E:JT}.
On the other hand, knowing the total number of DoFs in the global DoF vector, $D$,
the \objname{DMPlexSectionLoad} function chooses the partition map
$\chi_{\aset{J}}^{\aset{J}_P}$ \cref{E:chi_D_JP}, and then
identifies $\aset{J}_P^{m}$, and then $\aset{J}_P$ \cref{E:JP},
indexing the partitioned global DoF vector to be loaded.
The \objname{PetscSF} representing $\chi_{\aset{I}_T}^{\aset{I}_P}$ \cref{E:chi_IT_IP}
is used to broadcast the loaded offset data
$\anarray{LocOFF}_P^{m}$
as required in \cref{E:chi_JT_D_jT}, and finally,
the \objname{DMPlexSectionLoad} function constructs
a \objname{PetscSF} object representing
$\chi_{\aset{J}_T}^{\aset{J}_P}$ \cref{E:chi_JT_JP}
using the \objname{PetscSFCompose} function.

The \objname{load\_function} method of \objname{CheckpointFile} class
calls \objname{DMPlexSectionLoad} and
also loads the UFL element
if the function space interested has not been loaded yet,
when attempting to load an associated function;
see \cref{SS:function_impl}.

\subsection{Functions}\label{SS:function_impl}
Finally, we describe our implementation for saving and loading functions
following the discussions in \cref{SS:function} and \cref{SS:reconstruct}.
As discussed in \cref{SS:function},
a Firedrake function on a finite element function space is
represented by local/global DoF vector,
and it is represented by a PETSc \objname{Vec} object defined on the
associated local/global section,
i.e., local/global section along with the entity-local DoF ordering based on the cones
defines the map from DoFs in the local/global vector to members of the dual space.

We first consider saving the \objname{Vec}.
Once the global section has been saved with \objname{DMPlexSectionView}
as described in \cref{SS:function_space_impl},
we can use the \objname{DMPlexGlobalVectorView} function 
to save the \objname{Vec} representing the global vector $\ol{\anarray{VEC}}_P$
directly in association with the global section.
If more convenient, we can also use the \objname{DMPlexLocalVectorView} function
which first converts the local vector representation
$\ol{\anarray{LocVEC}}_T^{m}$ to $\ol{\anarray{VEC}}_P$ and then follows the same procedure as \objname{DMPlexGlobalVectorView}.
The \objname{save\_function} method of \objname{CheckpointFile} wraps
\objname{DMPlexGlobalVectorView}.

We now consider loading the saved \objname{Vec}.
Once \objname{DMPlexSectionLoad} is called
as described in \cref{SS:function_space_impl},
the local section is constructed
on the loaded \objname{DMPlex} topology.
The saved global DoF vector is partitioned using the map $\chi_{\aset{J}}^{\aset{J}_P}$ \cref{E:chi_D_JP}
and loaded in parallel as a \objname{Vec} object representing $\anarray{LocVEC}_P^{m}$
on each process.
We then create a \objname{Vec} object representing
$\anarray{LocVEC}_T^{m}$ according to the local section,
and fill in this \objname{Vec}
using the broadcasting formula \cref{E:VEC_bcast}
using the \objname{PetscSF} object representing
$\chi_{\aset{J}_T}^{\aset{J}_P}$ \cref{E:chi_JT_JP};
here, we again use the \objname{PetscSFBcast} function.

\objname{DMPlexLocalVectorLoad} function wraps the above series of operations.
\objname{DMPlexGlobalVectorLoad} function uses slightly modified \objname{PetscSF}
to directly broadcast the \objname{Vec} representing $\anarray{LocVEC}_P^{m}$ to
the \objname{Vec} representing the global DoF vector associated with the global section
on the loaded \objname{DMPlex} topology.
The \objname{load\_function} method of \objname{CheckpointFile} wraps
\objname{DMPlexGlobalVectorLoad}.

We use the same infrastructure for saving and loading
mesh coordinates, and wrap those operations in
\objname{DMPlexCoordinatesView} and \objname{DMPlexCoordinatesLoad}, respectively.
The \objname{save\_mesh} and the \objname{load\_mesh} methods
of \objname{CheckpointFile} calls these functions
automatically when saving and loading the mesh topology, respectively.

Finally, entities in a \objname{DMPlex} can be assigned \emph{labels}
indicating, e.g., boundary faces.
These labels are stored in \objname{DMLabel} objects.
These objects can also be saved and loaded in association with the \objname{DMPlex}
using a similar infrastructure;
they are implemented in \objname{DMPlexLabelsView} and \objname{DMPlexLabelsLoad},
and these functions are also called in
the \objname{save\_mesh} and the \objname{load\_mesh} methods
of \objname{CheckpointFile}.

\section{Orientations}\label{S:orientations}
We mentioned in \cref{SS:function} that,
given a finite element function space,
how the DoFs map to members of the dual space on each entity
is determined by the cone of that entity, and,
as the cones are preserved in the save-load cycle,
we can
reconstruct saved functions on the loaded mesh topology.
Thus, DoFs are laid out unambiguously on each physical mesh entity.
In the context of finite element analysis, however,
one maps a given cell to the reference cell and
works with the reference element with the reference DoFs
to assemble cell-local vectors and cell-local matrices.
Note that cones are also defined on the entities of the reference cell, and
the reference DoFs are laid out relative to the cones.
Mapping a physical mesh cell to the reference cell requires
transforming the DoFs on the physical mesh cell
to those on the reference cell,
and such transformation depends on
how we map the mesh cell to the reference cell.

Specifically, for each mesh entity mapped to a reference cell entity,
we determine the \emph{orientation}, represented by an integer,
by comparing the cone of the mesh entity
and that of the reference cell entity.
Then, we determine the correct DoF transformation according to the orientation.
For continuous and discontinuous Lagrange finite elements (the P, DP, Q, and DQ element families),
such transformation is given as a \emph{permutation} of the DoFs.

\input{paper_fig_orientation_fiat}

We illustrate this in \cref{Fi:orientation_fiat}
using P4 (CG4) finite element space
depicted in \cref{Fi:function_dofs_loaded};
specifically, we map the mesh cell with local number 1 on process 2
to the reference cell in a prescribed way
and construct permutations representing the DoF transformations.
A triangular mesh cell can be mapped to the reference cell
in $3!=6$ different ways, and
this example illustrates one of them.

For instance,
the mesh edge with local number 9 is mapped
to the FIAT reference edge labeled 4,
and the ordering of the cone of the former agrees with that of the latter
as indicated by the arrows pointing in the same direction;
this edge in the mesh is mapped
to the edge in the reference cell with orientation 0
in our definition.
The associated permutation is [0,1,2], i.e.,
the associated reference DoFs $\textrm{offset}+[0,1,2]$ ($\textrm{offset}=6$) is mapped to
the physical DoFs $\textrm{offset}+[0,1,2]$ ($\textrm{offset}=3$),
where $6+[0,1,2]$, for instance, represents $[6,7,8]$.

However, the mesh edge with the local number 10 is mapped
to the FIAT reference edge labeled 3,
and the ordering of the cone of the former disagrees with that of the latter as indicated by the arrows
pointing in the opposite directions;
this edge in the mesh is mapped to the edge in the reference cell with orientation 1
in our definition.
The associated permutation is [2,1,0], i.e.,
the associated reference DoFs $\textrm{offset}+[0,1,2]$ ($\textrm{offset}=3$) is mapped to
the physical DoFs $\textrm{offset}+[2,1,0]$ ($\textrm{offset}=6$).
There are only two possible orientations for each edge.
On the other hand, there are $3!=6$ possible orientations for the cell
and this example shows one of them;
the associated permutation is [2,0,1], i.e.,
the associated reference DoFs $\textrm{offset}+[0,1,2]$ ($\textrm{offset}=12$) is mapped to
the physical DoFs $\textrm{offset}+[2,0,1]$ ($\textrm{offset}=0$).

The surrounding packages of Firedrake,
FIAT~\cite{fiat2004} and FInAT~\cite{finat2017},
implement such permutations
for each entity for each possible orientation
for P, DP, Q, and DQ finite elements.
Traditionally, Firedrake uses DoF orderings based on the global numbers
of the mesh entities (following~\cite{rognes2010} S5.1)
as opposed to those based on the cones of the entities
as described in \cref{SS:function}, and,
except for P, DP, Q, and DQ spaces,
Firedrake still uses DoF orderings based on the global numbers;
this is unsafe for saving and loading functions
defined on the other finite element spaces that Firedrake supports
since global numbers are,
in general, not preserved in save-load cycles, unlike cones.
Thus, for the purpose of saving and loading functions,
we first project those functions,
if they are not defined on P, DP, Q, and DQ spaces,
onto appropriate DP/DQ spaces, save, load, and then
project them back onto the original function space.
Implementing DoF transformations
for all elements in FIAT and FInAT so that
Firedrake can save all functions natively is left to future work.

\section{Firedrake \objname{CheckpointFile} API}\label{S:firedrake_api}
\subsection{Basic usage}\label{SS:basic_usage}
The rather involved algorithm described above employs only information already encoded in the distributed in-memory representations of functions, function spaces and meshes employed by Firedrake. This enables a very simple high-level interface to this functionality.
\Cref{code:api_base} shows a basic usage of \objname{CheckpointFile}, where we save/load the mesh \objname{mesh} and the function \objname{f} to/from the file \objname{a.h5}. Importantly, this code will run without alteration on any number of processes.
Saving sequences of functions with the same name in the same function space is supported by passing an additional argument specifying an index in the function sequence.

\begin{figure}[!ht]
\centering
\begin{minted}[fontsize=\footnotesize]{python}
    from firedrake import *
    mesh = UnitSquareMesh(8, 8, name="my_mesh")
    V = FunctionSpace(mesh, "CG", 2)
    f = Function(V, name="my_func")
    with CheckpointFile("a.h5", "w") as ck:
        ck.save_mesh(mesh)  # optional
        ck.save_function(f)
    with CheckpointFile("a.h5", "r") as ck:
        mesh = ck.load_mesh("my_mesh")
        f = ck.load_function(mesh, "my_func")
\end{minted}
\captionof{listing}{A simple example of Firedrake code which saves and loads a mesh and a function from a file.}
\label{code:api_base}
\end{figure}

\section{Evaluation}\label{S:evaluation}
We ran several numerical experiments
to check correctness and test scalability.

\subsection{Correctness}\label{SS:correctness}
We tested the correctness of our implementation by saving functions and
loading them back, and verifying that the loaded functions are
DoF-wise the equal to the saved functions up to machine precision.
We used various process counts for saving and loading
and various finite element families, including
P, DP, Q, DQ, RTE/F, N1E/F, BDME/F, N2E/F, Q, DQ, RTCE/F, S, and DPC
with various degrees
on one-, two-, and three-dimensional meshes.
We also verified that our implementation works seamlessly for extrusion problems and time-dependent simulations.

\subsection{Performance}\label{SS:performance}
We ran our weak scalability tests for saving and loading meshes with associated functions
on ARCHER2, the UK National Supercomputing Service.
ARCHER2 uses the Lustre file system, with which one can save data
on multiple \emph{Object Storage Targets} (OSTs).
To store large files, one must \emph{stripe} them across multiple OSTs;
striping allows multiple processes to save to or load from the same file and increases the bandwidth.
One can set the number of OSTs over which the file is striped (stripe count)
and stripe size by passing the \texttt{--stripe-count/-c} and \texttt{--stripe-size/-S} option,
respectively, to the \texttt{lfs setstripe} command.
Next, we study the impact of these parameters and also two implementations of MPI libraries available on ARCHER2.

\subsubsection{Benchmark tests}\label{SSS:benchmark}
Before running our Firedrake tests,
we ran benchmark tests for HDF5 parallel output weak scalability~\cite{benchio} on ARCHER2.
See also~\cite{Jackson2011} for a study of I/O performance of parallel codes.
In the benchmark tests, each parallel process always saved 2.1 million double-precision numbers,
and we measured the output bandwidth in [GiB/s] for each case.

First, we studied the impact of the stripe count and stripe size parameters
for Lustre file system using both OFI and UCX implementations of MPI
using 64 ARCHER2 nodes with 128 processes per node.
We used three values, \texttt{1}, \texttt{4}, and \texttt{12}, for stripe count,
where 12 is the maximum, and, for each of them, used three values,
\texttt{4m}, \texttt{64m}, and \texttt{128m}, for stripe size, where \texttt{n} stands for MiB.
\Cref{Ta:eval_save_benchio_params_ofi,Ta:eval_save_benchio_params_ucx}
show results for OFI and UCX, respectively.
One can observe that UCX implementation shows higher bandwidths for most combinations of
stripe count and stripe size, and, for the UCX implementation, \texttt{-c=12} always outperformed the others.
Thus, in the rest of this section,
we use the UCX implementation of the MPI library and fix the stripe count to 12.

We then ran the weak scalability benchmark tests using 1, 8, 64, and 128 ARCHER2 nodes
with 128 processes per node
for stripe sizes \texttt{4m}, \texttt{64m}, and \texttt{128m},
and measured the bandwidth in GiB/s for each case.
Results are shown in \cref{Ta:eval_save_benchio_nodes}.
One can observe that the bandwidth saturates as the problem becomes larger around 10 GiB/s,
or writing to disk is \emph{bandwidth-limited},
which is what we expect according to the description given by ARCHER2.

\subsubsection{Firedrake tests}\label{SSS:firedrake}



\begin{table}
\centering 
\begin{subtable}[t]{0.48\textwidth}
  \centering
\begin{tabular}{rccc}
\toprule
  & \multicolumn{3}{c@{}}{Stripe size [MiB]}\\
  \cmidrule(l){2-4}
  Stripe count      & 4 & 64 & 128 \\
  \midrule
  1  & .24 & .83 & .82\\
  4  & 3.0 & 2.0 & 2.0\\
  12 & 5.3 & 1.6 & 1.6\\
  \bottomrule
\end{tabular}
\caption{OFI\label{Ta:eval_save_benchio_params_ofi}}
\end{subtable}
\begin{subtable}[t]{0.48\textwidth}
  \centering
\begin{tabular}{rccc}
  \toprule
  & \multicolumn{3}{c@{}}{Stripe size [MiB]} \\
    \cmidrule(l){2-4}
  Stripe count      & 4 & 64 & 128 \\
\midrule
1  & .42 & .78 & .27\\
4  & 1.8 & 2.4 & 1.1\\
12 & 9.3 & 9.5 & 3.2\\
\bottomrule
\end{tabular}
\caption{UCX\label{Ta:eval_save_benchio_params_ucx}}
\end{subtable}
\caption{
Weak scalability benchmark test for HDF5 parallel saving on Archer2
[\href{https://github.com/davidhenty/benchio}{benchio}]
run on 64 Archer2 nodes.
Three different stripe counts and
three different stripe sizes,
were explored for both
(a) the OFI implementations of the MPI library and
(b) the UCX implementations of the MPI library.
Each process saved about 2.1 million double-precision numbers
and the bandwidths in [GiB/s] were measured.
One can observe that the UCX implementation outperforms the OFI, and
(for UCX) a stripe count of twelve performs best independent of the
stripe size.
In the rest of this work we always use the UCX implementation with
stripe count set to twelve.\label{Ta:eval_save_benchio_params}}
\end{table}

\begin{table}
\centering 
\begin{tabular}{rccc}
  \toprule
  & \multicolumn{3}{c@{}}{Stripe size [MiB]} \\
  \cmidrule(l){2-4}
  Nodes & 4 & 64 & 128 \\
  \midrule
1    & 1.7 & 1.5 & 1.4 \\
8   & 7.7 & 6.2 & 4.5 \\
64  & 9.3 & 9.5 & 3.2 \\
128 & 8.6 & 9.5 & 9.7 \\
\bottomrule
\end{tabular}
\caption{
Weak scalability benchmark test for HDF5 parallel saving on Archer2 [\href{https://github.com/davidhenty/benchio}{benchio}]
run on 1, 8, 64, and 128 Archer2 nodes
for three different stripe sizes.
A fixed stripe count of twelve and the 
UCX implementation of the MPI library were used.
In this benchmark test each process saved about 2.1 million double-precision numbers
and the bandwidths in GiB/s were measured.
One can observe that the bandwidth saturates at around 10 GiB/s as indicated by the description given by ARCHER2.
The bandwidth observed with 64 nodes and 128 MiB stripe size was notably low and not in agreement with the other numbers;
we viewed this poor performance as a consequence of the very specific combination of the parameters, i.e., stripe size, node count, and size of data saved from each process, and regarded this as not of practical importance.
Given that the bandwidth was not very sensitive to the stripe size for the 64 node cases (excluding 128 MiB case) and the 128 node cases, we always use a stripe size of 128 MiB in the rest of the work.\label{Ta:eval_save_benchio_nodes}}
\end{table}

\begin{table}
\centering 
\begin{subtable}[t]{\textwidth}
  \centering
\begin{tabular}{rccccc}
  \toprule
  & \multicolumn{4}{c@{}}{Save time [s]} & \\
  \cmidrule(l){2-5}
  Nodes & \objname{Topology} & \objname{Labels} & \objname{Section} & \objname{Vec} & Bandwidth [GiB/s]\\
\midrule
1    & 5.5  & 3.6 & 2.9 & 1.9 & .51\\
8  & 22.  & 14. & 7.1 & 3.3 & 2.3\\
64 & 98. & 49. & 36. & 10. & 6.2\\
\bottomrule
\end{tabular}
\caption{64 processes per node\label{Ta:eval_save_firedrake_64ppn}}
\end{subtable}

\begin{subtable}[t]{\textwidth}
  \centering
\begin{tabular}{rccccc}
  \toprule
  & \multicolumn{4}{c@{}}{Save time [s]} & \\
  \cmidrule(l){2-5}
  Nodes & \objname{Topology} & \objname{Labels} & \objname{Section} & \objname{Vec} & Bandwidth [GiB/s]\\
1  & 4.4  & 2.1 & 1.4 & 2.0 & .49\\
8  & 25.  & 15. & 7.1 & 3.3 & 2.3\\
64 & 115. & 58. & 35. & 11. & 5.6\\
\bottomrule
\end{tabular}
\caption{128 processes per node\label{Ta:eval_save_firedrake_128ppn}}
\end{subtable}
\caption{
Weak scalability test for Firedrake HDF5 parallel saving on Archer2
run on 1, 8, and 64 Archer2 nodes, using
(a) 64 processes per node and (b) 128 processes per node.
The stripe count was set to twelve,
the stripe size was set to 128 MiB,
and the UCX implementation of the MPI library
was used throughout.
For each case, a tetrahedral mesh and a DP4 function were created
so that each process owned about 29 thousand cells and 1.0 million DoFs.
The largest problem with 64 nodes thus involved
a total of 0.24 billion cells and 8.3 billion DoFs.
Wall-clock time in seconds required for
\objname{DMPlexTopologyView} (\objname{Topology}),
\objname{DMPlexLabelsView} (\objname{Labels}),
\objname{DMPlexSectionView} (\objname{Section}) for the DP4 function space, and
\objname{DMPlexGlobalVectorView} (\objname{Vec}) for the DP4 function
was measured.
For \objname{DMPlexGlobalVectorView} we also computed the bandwidth for each case.
One can observe that, for both (a) and (b),
we achieved about 60\% of the target bandwidth.\label{Ta:eval_save_firedrake}}
\end{table}

We then performed a weak scalability test for Firedrake.
We prepared an unstructured tetrahedral mesh of a unit sphere with Gmsh~\cite{Gmsh}, loaded it as a Firedrake mesh, and refined it several times to obtain a \emph{base mesh}.
The base mesh contained 3,665,152 tetrahedral cells.
This problem size is comparable to those considered
in the previous section, so we further consistently use the UCX implementation of the MPI library and fix stripe count to 12 and stripe size to \texttt{128m}.

We then created a series of meshes by further refining the base mesh;
we refer to the number of times the base mesh was refined,
in a manner that doubles the number of cells in each direction in each cell,
as \emph{refinement level}, and denote it by $r$.
For our weak scalability test for saving,
we used meshes of $r = 0,1,2$,
and, on each, constructed a DP4 function space and a function on that function space.

To save functions defined on meshes of $r = 0,1,2$,
we used 1, 8, and 64 ARCHER2 nodes, respectively,
so that each process handled about 29 thousand cells and 1.0 million DoFs if 128 processes were used per node.
The largest problem with 64 nodes thus involved a total of 234,569,728 cells and 8,209,940,480 DoFs.
For each case, we measured wall-clock times in seconds
required for \objname{DMPlexTopologyView} for saving the mesh topology, \objname{DMPlexLabelsView} for saving the labels,
\objname{DMPlexSectionView} for saving the DP4 function space, and \objname{DMPlexGlobalVectorView} for saving the DP4 function.
For \objname{DMPlexGlobalVectorView}, we also computed the bandwidth in GiB/s.
\Cref{Ta:eval_save_firedrake} shows results for 64 and 128 processes per node.

One can observe that for both cases, we were able to achieve about 60\% of the target bandwidth on ARCHER2, 10 GiB/s, using 64 nodes
for saving the function data.
One can also see that saving mesh topology and labels took
much more time than saving function data;
this is presumably due to saving many integer arrays of various sizes when saving topology and labels,
while the local \objname{Vec} representing the function data is a large contiguous array.
Another possible reason is that we optimized \texttt{--stripe-size} primarily for function data;
the rationale for that is, in many applications such as time-series simulations and checkpointing in inverse problems,
we save and load function data much more often than mesh topology data and labels.




\begin{table}
\centering 
\begin{subtable}[t]{\textwidth}
  \centering
\begin{tabular}{rccccc}
  \toprule
  & \multicolumn{5}{c@{}}{Load time [s]} \\
  \cmidrule(l){2-6}
  Nodes & \objname{Topology} & \objname{Distribute} & \objname{Labels} & \objname{Section} & \objname{Vec}\\
  \midrule
  1   &  1.5 & 15. &  1.1 & 1.4 & .46\\
  8  &  43. & 22. &  22. & 9.8 & .53\\
  64 & 382. & 39. & 209. & 171. & 1.3\\
  \bottomrule
\end{tabular}
\caption{64 processes per node\label{Ta:eval_load_firedrake_64ppn}}
\end{subtable}

\begin{subtable}[t]{\textwidth}
  \centering
\begin{tabular}{rccccc}
  \toprule
  & \multicolumn{5}{c@{}}{Load time [s]} \\
  \cmidrule(l){2-6}
  Nodes & \objname{Topology} & \objname{Distribute} & \objname{Labels} & \objname{Section} & \objname{Vec}\\
  \midrule
  1   &  1.3 & .71 &  1.0 &  .82 & .41\\
  8  &  41. & 1.5 &  21. &  14. & .48\\
  64 & 378. & 13. & 203. & 168. & 1.8\\
  \bottomrule
\end{tabular}
\caption{128 processes per node\label{Ta:eval_load_firedrake_128ppn}}
\end{subtable}
\caption{
Weak scalability test for Firedrake HDF5 parallel loading on Archer2
run on 1, 8, and 64 Archer2 nodes, using
(a) 64 processes per node and (b) 128 processes per node.
The loaded meshes were redistributed using \objname{DMPlexDistribute} function with ParMETIS backend.
The stripe count was set to twelve,
the stripe size was set to 128 MiB,
and the UCX implementation of the MPI library
was used throughout.
For each case, the tetrahedral mesh and the DP4 function
that were saved in the corresponding problem in the saving tests were loaded.
Wall-clock time in seconds required for 
\objname{DMPlexTopologyLoad} (\objname{Topology}),
\objname{DMPlexDistribute} (\objname{Distribute}),
\objname{DMPlexLabelsLoad} (\objname{Labels}),
\objname{DMPlexSectionLoad} (\objname{Section}) for the DP4 function space, and
\objname{DMPlexGlobalVectorLoad} (\objname{Vec}) for the DP4 function
was measured.
One can observe similar tendency as in the saving tests
of requiring more time as the problem becomes larger;
see also \cref{Ta:eval_save_firedrake}.
\label{Ta:eval_load_firedrake}}
\end{table}
\begin{table}
\centering 
\begin{subtable}[t]{\textwidth}
  \centering
\begin{tabular}{rccccc}
  \toprule
  & \multicolumn{5}{c@{}}{Load time [s]} \\
  \cmidrule(l){2-6}
  Nodes & \objname{Topology} & \objname{Distribute} & \objname{Labels} & \objname{Section} & \objname{Vec}\\
  \midrule
1   & 1.5  & 3.2 & 1.1 & 1.4 & .48\\
8  & 43.  & 18. & 21. & 10. & .55\\
64 & 377. & 124.& 207.& 168.&  1.4\\
\bottomrule
\end{tabular}
\caption{64 processes per node\label{Ta:eval_load_firedrake_dist_64ppn}}
\end{subtable}

\begin{subtable}[t]{\textwidth}
  \centering
\begin{tabular}{rccccc}
  \toprule
  & \multicolumn{5}{c@{}}{Load time [s]} \\
  \cmidrule(l){2-6}
  Nodes & \objname{Topology} & \objname{Distribute} & \objname{Labels} & \objname{Section} & \objname{Vec}\\
  \midrule
1   & 1.3 & 1.9 & 1.0 & 1.0 & .51\\
8  & 42. & 15. & 19. & 15. & .58\\
64 & 360.& 115.& 200.& 163.& 1.3\\
\bottomrule
\end{tabular}
\caption{128 processes per node\label{Ta:eval_load_firedrake_dist_128ppn}}
\end{subtable}
\caption{
Weak scalability test for Firedrake HDF5 parallel loading on Archer2
run on 1, 8, and 64 Archer2 nodes, using
(a) 64 processes per node and (b) 128 processes per node.
The loaded meshes were redistributed so that they have the exact same distribution as the meshes before saving.
The stripe count was set to twelve,
the stripe size was set to 128 MiB,
and the UCX implementation of the MPI library
was used throughout.
For each case, the tetrahedral mesh and the DP4 function
that were saved in the corresponding problem in the saving tests were loaded.
Wall-clock time in seconds required for 
\objname{DMPlexTopologyLoad} (\objname{Topology}),
\objname{DMPlexDistribute} (\objname{Distribute}),
\objname{DMPlexLabelsLoad} (\objname{Labels}),
\objname{DMPlexSectionLoad} (\objname{Section}) for the DP4 function space, and
\objname{DMPlexGlobalVectorLoad} (\objname{Vec}) for the DP4 function
was measured.
One can observe similar tendency as in the saving tests
of requiring more time as the problem becomes larger;
see also \cref{Ta:eval_save_firedrake}.\label{Ta:eval_load_firedrake_dist}}
\end{table}

Further, we loaded each saved mesh-function pair using the same number of nodes.
For each case, we measured wall-clock times in seconds
required for \objname{DMPlexTopologyLoad}, \objname{DMPlexLabelsLoad}, \objname{DMPlexSectionLoad}, and \objname{DMPlexGlobalVectorLoad}.
We ran this experiment twice; the first time using \objname{DMPlexDistribute} function with the ParMETIS~\cite{parmetis} backend to redistribute the loaded meshes, and the second time loading the mesh distributions directly to recover the exact same parallel mesh distributions as before saving.
We used 64 and 128 processes per node, as
using a smaller number of processes per node relaxes some memory overhead, e.g., that of partitioners.
\Cref{Ta:eval_load_firedrake} shows results using \objname{DMPlexDistribute}.
\Cref{Ta:eval_load_firedrake_dist}
shows results when the saved parallel distributions were recovered in the loading time.
As for saving, loading the mesh topology and labels took much more time than loading function data, and
recovering the saved mesh distribution took much longer than redistributing using \objname{DMPlexDistribute},
presumably, because it involved loading additional arrays representing the distribution of the mesh before saving.
Again, this might have been partially caused by the fact that we optimized \texttt{--stripe-size} for saving and loading function data.

\section{Conclusions}\label{S:conclusion}
We introduced an efficient N-to-M checkpointing algorithm for finite element simulations, which allows saving and loading
function spaces and functions along with the mesh using a different number of parallel processes.
We showed that to overcome the challenge of reconstructing function spaces and functions defined on the mesh before saving on an arbitrarily distributed loaded mesh,
we need to identify appropriate sets of entities and DoFs, and construct maps between them.
These maps can be seen as star forest graphs,
where global numbers that we computed on the mesh before saving play an important role.
This concept was then seamlessly translated to PETSc using its \objname{PetscSF} class.
By this means, we implemented efficient N-to-M mesh I/O capabilities in PETSc,
which were then wrapped in the new \objname{CheckpointFile} class in Firedrake.
We verified the correctness of the new I/O feature by saving and loading various finite element function spaces and functions along with the meshes in Firedrake using a different number of parallel processes for saving and loading, which included extrusion and timestepping problems.
We also observed good I/O performance on ARCHER2 for problems with up to 8 billion DoFs using 64 nodes with 128 processes per node.

In Firedrake, by default, if the process count for loading is the same as for saving,
the loaded mesh will have the exact same distribution as the saved mesh, and
the global numbers computed on the loaded mesh will be identical to those computed on the mesh before saving.
This, however, is not the case in general, especially when we use a different process count for loading,
and the global numbers of entities computed on the loaded mesh will differ from those computed on the mesh before saving.
This means that if we save the loaded mesh back to the file,
the mesh is viewed as a new mesh, and function spaces and functions defined on the loaded mesh must be saved under this new mesh.
This issue will be resolved in PETSc and thus in Firedrake once we make the loaded mesh inherit global numbers from the saved mesh,
which we have yet to address in the future work.

\section*{Acknowledgment}
This work was funded under the embedded CSE programme of the ARCHER2
UK National Supercomputing Service [http://www.archer2.ac.uk] and the Engineering and Physical Sciences Research Council [grant numbers EP/R029423/1, EP/W029731/1, EP/W026066/1], the National Science Foundation [NSF CSSI: 1931524], the Department of Energy [DOE Contract DE-AC02-06CH11357],
and the Swiss Platform for Advanced Scientific Computing (PASC) [projects ``Salvus'' (2017-2020) and ``Bayesian Waveform Inversion'' (since 2021)].
For the purpose of open access, the authors have applied a creative commons attribution (CC BY) licence to any author accepted manuscript version arising.

\section*{Code availability}
The exact version of Firedrake used, along with all of the scripts employed in the experiments presented in this paper has been archived on Zenodo~\cite{firedrake-zenodo}.

\section*{Remark}
This is a more detailed version of an ARCHER2 technical report~\cite{ecse_io}.

\appendix
\section{List of symbols}\label{S:symbols}
\nomenclature[01]{$\ol{\cdot}$}{Over-left-arrow for sets, arrays, and other quantities on the mesh before saving; \emph{no over-left-arrow} for those on the loaded mesh}
\nomenclature[04]{$T$}{Subscript for data directly associated with the mesh topology; $T$ stands for \emph{topology}}
\nomenclature[05]{$P$}{Subscript for partitioned data; $P$ stands for \emph{partition}}
\nomenclature[06]{$N$}{Number of process used in the saving session}
\nomenclature[07]{$M$}{Number of process used in the loading session ($\neq N$ in general)}
\nomenclature[08]{$E$}{Total number of entities}
\nomenclature[10]{$\ol{E}_T^{n}$}{Number of entities visible to process $n\in\{0,\dots,N\minus 1\}$ on the parallel mesh topology before saving}
\nomenclature[14]{$E_T^{m}$}{Number of entities visible to process $m\in \{0,\dots,M\minus 1\}$ on the loaded parallel mesh topology}
\nomenclature[16]{$E_P^{m}$}{Size of the $m$-th partition of the array $[0:E)$ determined when loading the function space}
\nomenclature[17]{$D$}{Total number of DoFs}
\nomenclature[18]{$\ol{D}_T^{n}$}{Number of DoFs visible to process $n\in\{0,\dots,N\minus 1\}$ on the parallel mesh topology before saving}
\nomenclature[19]{$D_T^{m}$}{Number of DoFs visible to process $m\in \{0,\dots,M\minus 1\}$ on the loaded parallel mesh topology}
\nomenclature[20]{$D_P^{m}$}{Size of the $m$-th partition of the array $[0:D)$ determined when loading the function}
\nomenclature[21]{$\aset{I}$}{Set of global numbers of entities $\{0,\cdots,E\minus 1\}$}
\nomenclature[22]{$\ol{\aset{I}}_T^{n}$}{Set of local numbers of entities visible to process $n\in\{0,\dots,N\minus 1\}$ on the parallel mesh topology before saving, $\{0,\dots,\ol{E}_T^{n}\minus 1\}$}
\nomenclature[23]{$\aset{I}_T^{m}$}{Set of local numbers of entities visible to process $m\in \{0,\dots,M\minus 1\}$ on the loaded parallel mesh topology, $\{0,\dots,E_T^{m}\minus 1\}$}
\nomenclature[24]{$\aset{I}_P^{m}$}{Set of local indices, $\{0,\dots,E_P^{m}\minus 1\}$}
\nomenclature[25]{$\aset{J}$}{Set of indices into the global vector, $\{0,\cdots,D\minus 1\}$}
\nomenclature[26]{$\ol{\aset{J}}_T^{n}$}{Set of indices into the local vector defined on the parallel mesh topology before saving, $\{0,\dots,\ol{D}_T^{n}\minus 1\}$}
\nomenclature[27]{$\aset{J}_T^{m}$}{Set of indices into the local vector defined on the loaded parallel mesh topology, $\{0,\dots,D_T^{m}\minus 1\}$}
\nomenclature[28]{$\aset{J}_P^{m}$}{Set of indices, $\{0,\dots,D_P^{m}\minus 1\}$}
\nomenclature[33]{$\aset{L}_P$}{Set of tuples of a process ID and a local index, from which a bijective map exists to the set of global numbers representing mesh entities, $\aset{I}$}
\nomenclature[40]{$\ol{\anarray{LocVEC}}_T^{n}$}{Local vector defined on the parallel mesh topology before saving indexed by $\ol{J}_T^{n}$}
\nomenclature[41]{$\ol{\anarray{LocOFF}}_T^{n}$}{Local array of offsets in $\ol{\anarray{LocVEC}}_T^{n}$ defined on the parallel mesh topology before saving indexed by $\ol{I}_T^{n}$}
\nomenclature[42]{$\ol{\anarray{LocDOF}}_T^{n}$}{Local array of numbers of DoFs defined on the parallel mesh topology before saving indexed by $\ol{I}_T^{n}$}
\nomenclature[43]{$\ol{\anarray{LocG}}_T^{n}$}{Local array of global numbers defined on the parallel mesh topology before saving indexed by $\ol{I}_T^{n}$}
\nomenclature[50]{$\ol{\anarray{VEC}}_P$}{Partitioned global vector of total size $D$ made from $\ol{\anarray{LocVEC}}_T^{n}$, $n\in\{0,\dots,N\minus 1\}$, when saving}
\nomenclature[51]{$\ol{\anarray{OFF}}_P$}{Partitioned global array of total size $E$ of offsets in $\ol{\anarray{VEC}}_P$ made from $\ol{\anarray{LocOFF}}_T^{n}$, $n\in\{0,\dots,N\minus 1\}$, when saving}
\nomenclature[52]{$\ol{\anarray{DOF}}_P$}{Partitioned global array of total size $E$ of numbers of DoFs made from $\ol{\anarray{LocDOF}}_T^{n}$, $n\in\{0,\dots,N\minus 1\}$, when saving}
\nomenclature[53]{$\ol{\anarray{G}}_P$}{Partitioned global array of total size $E$ of global numbers made from $\ol{\anarray{LocG}}_T^{n}$, $n\in\{0,\dots,N\minus 1\}$, when saving}
\nomenclature[60]{$\anarray{VEC}_P$}{Partitioned global vector of total size $D$; partitioned according to $D_P^{m}$, $m\in \{0,\dots,M\minus 1\}$, when loading}
\nomenclature[61]{$\anarray{OFF}_P$}{Partitioned global array of total size $E$ of offsets in $\anarray{VEC}_P$; partitioned according to $E_P^{m}$, $m\in \{0,\dots,M\minus 1\}$, when loading}
\nomenclature[62]{$\anarray{DOF}_P$}{Partitioned global array of total size $E$ of numbers of DoFs; partitioned according to $E_P^{m}$, $m\in \{0,\dots,M\minus 1\}$, when loading}
\nomenclature[63]{$\anarray{G}_P$}{Partitioned global array of size $E$ of global numbers; partitioned according to $E_P^{m}$, $m\in \{0,\dots,M\minus 1\}$, when loading}
\nomenclature[70]{$\anarray{LocVEC}_P^{m}$}{Locally owned part of $\anarray{VEC}_P$ indexed by $\aset{J}_P^{m}$}
\nomenclature[71]{$\anarray{LocOFF}_P^{m}$}{Locally owned part of $\anarray{OFF}_P$ indexed by $\aset{I}_P^{m}$}
\nomenclature[72]{$\anarray{LocDOF}_P^{m}$}{Locally owned part of $\anarray{DOF}_P$ indexed by $\aset{I}_P^{m}$}
\nomenclature[73]{$\anarray{LocG}_P^{m}$}{Locally owned part of $\anarray{G}_P$ indexed by $\aset{I}_P^{m}$}
\nomenclature[80]{$\anarray{LocVEC}_T^{m}$}{Local vector defined on the loaded parallel mesh topology indexed by $J_T^{m}$}
\nomenclature[81]{$\anarray{LocOFF}_T^{m}$}{Local array of offsets in $\anarray{LocVEC}_T^{m}$ defined on the loaded parallel mesh topology indexed by $I_T^{m}$}
\nomenclature[82]{$\anarray{LocDOF}_T^{m}$}{Local array of numbers of DoFs defined on the loaded parallel mesh topology indexed by $I_T^{m}$}
\nomenclature[83]{$\anarray{LocG}_T^{m}$}{Local array of global numbers defined on the loaded parallel mesh topology indexed by $I_T^{m}$}
\nomenclature[99]{$\chi_\aset{A}^\aset{B}$}{Map from set $A$ to set $B$}
\printnomenclature[60pt]

\section{Loading and redistributing mesh topology}\label{S:loaded_mesh_details}
In this section we give details of the intermediate steps to reconstruct the mesh topology
on $M$ parallel processes upon loading.
Specifically, we describe how we obtain
the map $\chi_{\aset{I}_T}^{\aset{L}_P}$ \cref{E:chi_IT_LP}
using $\chi_{\aset{I}}^{\aset{L}_P}$ \cref{E:chi_E_LP},
with $\aset{I}_T$ \cref{E:IT} and $\aset{L}_P$ \cref{E:LP} defined in \cref{SSS:loaded_mesh}.

As mentioned in \cref{SSS:loaded_mesh}, 
reconstruction of the mesh topology on the loading processes is in general a three-step process:
\begin{enumerate}
    \item construct the mesh topology with no cell overlaps using a naive partition of the saved mesh topology data, i.e., the cones of the entities,
    \item redistribute it using a mesh partitioner, and,
    \item add overlaps across process boundaries if necessary.
\end{enumerate}
In each step we obtain different parallel representation of the same mesh topology.
Let $E_{T^{00}}^{m}$ and $E_{T^{0}}^{m}$ denote the number of entities visible to process $m\in \{0,\dots,M\minus 1\}$
in the mesh topologies obtained in Step 1 and Step 2, respectively.
Entities of these mesh topologies
are arbitrarily labeled locally with local numbers
drawn from sets
$\aset{I}_{T^{00}}^{m}:=\{0,\dots,E_{T^{00}}^{m}\minus 1\}$ and
$\aset{I}_{T^{0}}^{m}:=\{0,\dots,E_{T^{0}}^{m}\minus 1\}$.
To represent these maps monolithically over all parallel processes,
we define union sets, $\aset{I}_{T^{00}}$ and $\aset{I}_{T^{0}}$, as:
\begin{subequations}
\begin{align}
\aset{I}_{T^{00}}&:=
\bigcup\limits_{m\in\{0,\dots,M\minus 1\}}\{m\}\times\aset{I}_{T^{00}}^{m},\label{E:IT00}\\
\aset{I}_{T^{0}}&:=
\bigcup\limits_{m\in\{0,\dots,M\minus 1\}}\{m\}\times\aset{I}_{T^{0}}^{m}.\label{E:IT0}
\end{align}
\end{subequations}
Recall from \cref{SSS:loaded_mesh} that, for the final mesh topology constructed in Step 3,
$E_T^{m}$ denotes the number of entities visible to process $m\in \{0,\dots,M\minus 1\}$,
$\aset{I}_T^{m}:=\{0,\dots,E_T^{m}\minus 1\}$ denotes the set of local numbers, and
$\aset{I}_T$ \cref{E:IT} denotes the union set.

As the cones of the entities are written in terms of global numbers, $\aset{I}$, upon loading,
in Step 1, using the partition formula
$\chi_{\aset{I}}^{\aset{L}_P}$ \cref{E:chi_E_LP},
one can construct a map:
\begin{align}
\chi_{\aset{I}_{T^{00}}}^{\aset{L}_P}:\aset{I}_{T^{00}}\rightarrow\aset{L}_P.
\label{E:chi_IT00_LP}
\end{align}
On the other hand, in Step 2 and Step 3, one can construct surjective maps:
\begin{subequations}
\begin{align}
\chi_{\aset{I}_{T^{0}}}^{\aset{I}_{T^{00}}}&:\aset{I}_{T^{0}}\rightarrow\aset{I}_{T^{00}},
\label{E:chi_IT0_IT00}\\
\chi_{\aset{I}_T}^{\aset{I}_{T^{0}}}&:\aset{I}_T\rightarrow\aset{I}_{T^{0}}.
\label{E:chi_IT_IT0}
\end{align}
\end{subequations}
These maps can be viewed as star forests; see \cref{SSS:loaded_mesh}.
One can then compose
$\chi_{\aset{I}_{T^{00}}}^{\aset{L}_P}$ \cref{E:chi_IT00_LP},
$\chi_{\aset{I}_{T^{0}}}^{\aset{I}_{T^{00}}}$ \cref{E:chi_IT0_IT00}, and
$\chi_{\aset{I}_T}^{\aset{I}_{T^{0}}}$ \cref{E:chi_IT_IT0},
to obtain $\chi_{\aset{I}_T}^{\aset{L}_P}$ \cref{E:chi_IT_LP} as:
\begin{align}
\chi_{\aset{I}_T}^{\aset{L}_P}:=
\chi_{\aset{I}_{T^{00}}}^{\aset{L}_P}\circ
\chi_{\aset{I}_{T^{0}}}^{\aset{I}_{T^{00}}}\circ
\chi_{\aset{I}_T}^{\aset{I}_{T^{0}}}.
\label{E:chi_IT_LP_compose}
\end{align}

In practice, Steps 1, 2, and 3 are performed by
\objname{DMPlexTopologyLoad},
\objname{DMPlexDistribute}, and
\objname{DMPlexDistributeOverlap},
and these PETSc functions construct and return \objname{PetscSF}s representing, respectively,
$\chi_{\aset{I}_{T^{00}}}^{\aset{L}_P}$ \cref{E:chi_IT00_LP},
$\chi_{\aset{I}_{T^{0}}}^{\aset{I}_{T^{00}}}$ \cref{E:chi_IT0_IT00}, and
$\chi_{\aset{I}_T}^{\aset{I}_{T^{0}}}$ \cref{E:chi_IT_IT0}.
Finally, the map compositions in \cref{E:chi_IT_LP_compose} are performed using \objname{PetscSFCompose} function.

\bibliographystyle{siamplain}
\bibliography{paper}

\end{document}